\documentclass[a4paper,11pt]{article}
\pdfoutput=1 
\usepackage{jheppub} 
\usepackage{epsfig}
\usepackage{caption}
\usepackage{subcaption}
\usepackage{color,xcolor,soul,bm}
\usepackage{float}
\usepackage{amsfonts}
\usepackage{url}
\usepackage{slashed}
\usepackage{accents}
\usepackage{lipsum}
\usepackage{mathtools}
\usepackage{physics}

\usepackage[normalem]{ulem}

\hypersetup{
  bookmarks=true,         % show bookmarks bar?
  unicode=false,          % non-Latin characters in Acrobat?s bookmarks
  pdftoolbar=true,        % show Acrobat?s toolbar?
 pdfmenubar=true,        % show Acrobat?s menu?
 pdffitwindow=true,     % window fit to page when opened
 pdfstartview={FitH},    % fits the width of the page to the window
 pdfsubject={Dark Matter},   % subject of the document
 pdfnewwindow=true,      % links in new window
 pdfcreator={RevTeX},
 colorlinks=true,       % false: boxed links; true: colored links
 linkcolor=blue,          % color of internal links
 citecolor=magenta,        % color of links to bibliography
 filecolor=black,      % color of file links
 urlcolor=blue,           % color of external links
  }

\newcommand{\Tns}{T_\mathrm{NS}}
\newcommand{\Mns}{M_\mathrm{NS}}
\newcommand{\Rns}{R_\mathrm{NS}}
\newcommand{\sv}{\expval{\sigma v}}

\title{Probing freeze-in dark matter using Bose-Einstein condensate in neutron star}

\author[1,2]{Deep Ghosh}
\author[1,2]{\& Anirban Das}

\affiliation[1]{Saha Institute of Nuclear Physics,
Sector-I, Block-AF, Bidhannagar, Kolkata 700064, India.}
\affiliation[2]{Homi Bhabha National Institute, Training School Complex, Anushaktinagar, Mumbai 400094, India.}

\emailAdd{matrideb1@gmail.com}
\emailAdd{anirbandas.21@protonmail.com}

\abstract{Neutron star (NS) is one of the most promising astrophysical targets to probe non-gravitational interaction of dark matter (DM) with visible matter. Their compactness makes them an ideal object which can capture particle DM efficiently over its lifetime using the DM-nucleon scattering cross-section. If DM particles are bosonic, then the captured DM population may form a \emph{Bose-Einstein condensate} at the center of the NS, increasing the DM density significantly. In this work, we study the phenomenology of such scenario with enhanced DM annihilation rate due to the increased density in a condensate. The enhanced DM annihilation makes the NS surface `\textit{hotter}' than in the standard cooling scenario. We show that the annihilation rate is enhanced by a factor of  $\mathcal{O}(10^{15}-10^{20})$ if DM forms a condensate without large self-interaction, and DM with \emph{freeze-in} value annihilation cross-section can heat up the NS to higher temperatures, bringing it within the reach of James Webb Space Telescope. It also allows us to probe DM-nucleon scattering cross section within the \emph{neutrino fog} regime which will complement the terrestrial direct detection searches. Moreover, the enhanced annihilation from the condensate changes the lower limits on s-wave DM annihilation cross-section for capture-annihilation equilibrium and the formation of a black hole inside the NS. Finally, we show an example of a scalar DM model where such small annihilation and DM-nucleon scattering cross sections can generically arise.

}
\begin{document} 
\maketitle
\flushbottom  
\section{Introduction}
New generation telescopes and gravitational wave detectors have ushered in a new age of astrophysical observation in the last few years and will continue to revolutionize the area in the coming years in various ways. One of these is the observation of old neutron stars (NS) at the  galactic center\,\cite{Chennamangalam:2013zja,2018JKAS...51..165K,Zhang:2014kva,Rajwade:2016cto}, globular clusters\,\cite{Camilo:2005aa,Padmanabh:2024bsz,SKAPulsarScienceWorkingGroup:2025cun}, and also in the solar system neighborhood.
A plethora of studies have shown that NSs can be used as a natural laboratory to probe non-gravitational interactions of dark matter (DM) with the Standard Model (SM)\,\cite{Goldman:1989nd,Kouvaris:2007ay,Bramante:2013hn,Bell:2013xk,McDermott:2011jp,deLavallaz:2010wp,Acevedo:2019agu,Dasgupta:2020mqg,Leane:2021ihh,Bhattacharya:2023stq,Baryakhtar:2017dbj,Guver:2012ba,Raj:2017wrv,Bramante:2021dyx,Bose:2021yhz,Maity:2021fxw,Das:2024thc,Dutta:2024vzw,Ema:2024wqr,Robles:2025dlv,Dey:2025atz,Bhattacharya:2025xko,Liu:2025qco,Pospelov:2026dnb,Sarkar:2026fge}. The observational effects of DM-admixed NS with self-interaction have also been studied extensively \cite{Ellis:2018bkr,Gresham:2018rqo,Karkevandi:2021ygv,Das:2020ecp,Shakeri:2022dwg,Giangrandi:2022wht}. For a comprehensive review on these topics, see Ref.\,\cite{Bramante:2023djs}. On the other hand, the multi-tonne scale direct detection experiments, such as PandaX-4T, XENONnT, and LUX-ZEPLIN, have recently run into the solar \emph{neutrino fog} at the small cross-section regime that will slow down their further progress in the near future\,\cite{PandaX:2024muv,XENON:2024ijk, LZ:2025igz}. To probe lighter sub-GeV mass DM in laboratory, multiple experiments with lower thresholds are running and have constrained parts of the theory space\,\cite{CRESST:2022lqw,SuperCDMS:2020ymb,CDEX:2021cll,Das:2022srn,Das:2024jdz,Griffin:2024jec,TESSERACT:2025tfw,Hu:2025dsv,Schwemmbauer:2025evp}. However, their current sensitivity is relatively weaker. In this context, therefore the use of NSs to probe DM-nucleon interaction is both necessary and timely as a complementary channel.

In the presence of the DM-neutron interaction, DM particles undergo scattering and get captured inside the NS from the environment. The captured DM particles \emph{thermalize} inside the NS at the core temperature $\Tns$ and eventually form a \emph{dark core}.
If the DM particles annihilate into SM particles that get trapped inside the NS, it will act as a new source of heating inside the NS. This heating effect is relatively easy to observe in old NSs with age $\gtrsim 10^8$ years, for which cooling due to neutrino and photon emission becomes sub-dominant within the minimal cooling scenario \cite{Page:2004fy,Yakovlev:2003qy} . A conservative upper limit on the surface temperature of an old neutron star $T_s \leq 42000 {~\rm K}$ (3.6 eV), has been inferred from the observation of the slow-spinning, isolated $~3\times 10^8$ year-old pulsar PSR J2144–3933 using the Hubble Space Telescope \cite{2019ApJ...874..175G}. In addition, old NSs, e.g. PSR J0437-4715 \cite{Kargaltsev:2003eb,2012ApJ...746....6D}, PSR J0108-1431 \cite{Mignani:2008jr}, were observed to have a surface temperature around $10^5$\,K. In contrast, within the minimal cooling scenario, the predicted surface temperature of such an old NS is approximately $1$\,K \cite{Page:2004fy,Yakovlev:2003qy,Dutta:2024vzw}, well below the observational bounds, thereby hinting toward additional heating sources \cite{1995ApJ...442..749R,Fujiwara:2023tmr,Baryakhtar:2017dbj}, such as DM annihilation \cite{Kouvaris:2007ay}. The infrared sensitivity ($T_s\gtrsim  \mathcal{O}(10^3)$ K) of the James Webb Space Telescope (JWST)\,\cite{Gardner:2006ky} now provides an opportunity to probe the late-time thermal emission of old neutron stars, offering a potential test of the DM annihilation–induced heating hypothesis\,\cite{Chatterjee:2022dhp,Raj:2024kjq}.
 \begin{figure}[!t]
\includegraphics[width=\textwidth]
 {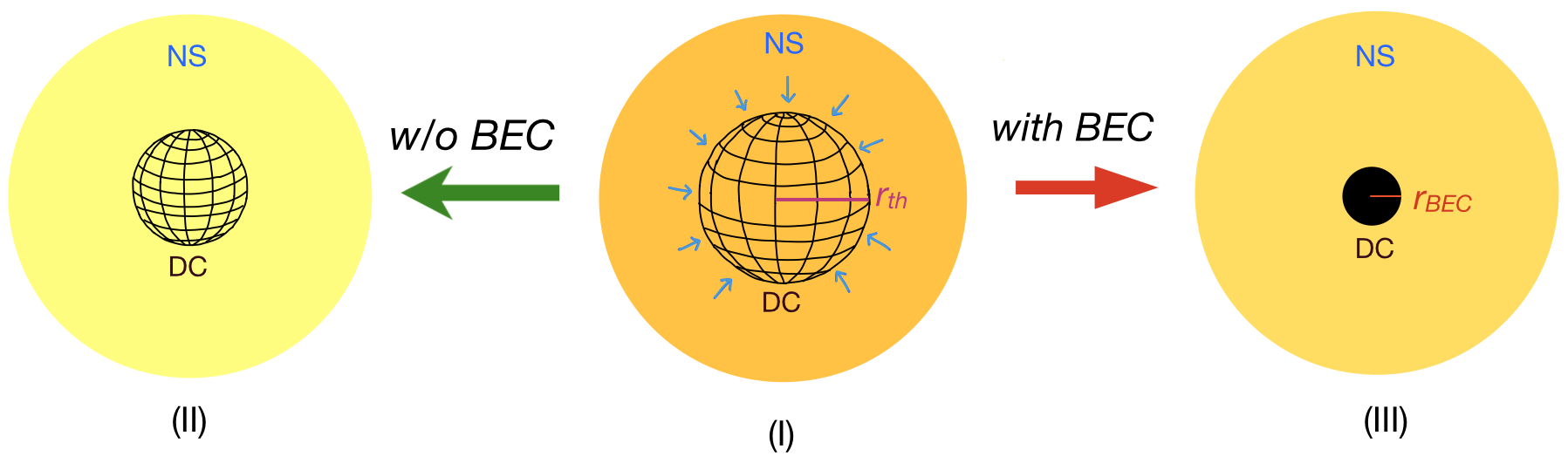}
 \caption{\textit{Schematic representation (not to scale) of dark matter (DM) dynamics inside a neutron star (NS)}: \textbf{I.} A dark core (DC) is formed  inside a NS after thermalization with a thermal radius $r_{\rm th}$. \textbf{II.} As the NS cools, $r_{th}$ decreases as $r_{\rm th}\propto\sqrt{\Tns}$. \textbf{III.} If DM particles form Bose-Einstein condensate (BEC), the DC suddenly shrinks to a very small region with a radius, $r_{\rm BEC} (\ll r_{\rm th})$. As a consequence, the annihilation rate inside the NS is enhanced by several orders of magnitude. Thus, the surface of an old NS with a BEC dark core is `\textit{hotter}' even for very small annihilation cross-sections, compared to the case without a BEC. A darker yellow shade represents higher NS surface temperature.}
 \label{fig:illus}
 \end{figure}

If the DM annihilation is efficient enough, the capture and the annihilation processes can reach an equilibrium in the core of the NS.
This capture-annihilation (CA) equilibrium timescale is controlled by thermally averaged annihilation cross-section $\expval{\sigma v}$ and the DM-neutron elastic scattering cross-section $\sigma_{\chi n}$. If the CA equilibrium is established within the age of a NS, the heating rate due to DM annihilation is at its maximum and effectively becomes a function of $\sigma_{\chi n}$ and the local DM density. Therefore, one can probe $\sigma_{\chi n}$ in a broad range of DM mass using observations of JWST. The required DM annihilation cross-section to achieve the CA equilibrium in $10^9$ years was previously calculated as $\expval{\sigma v}\gtrsim 10^{-54}{~\rm cm^3 ~s^{-1}}$, assuming a $10^3$\,K NS and the s-wave annihilation of 1\,GeV mass DM\,\cite{Garani:2020wge}.

For non-annihilating or very feebly annihilating DM scenarios, the captured DM population inside the NS can cross a critical value, making the dark core collapse. As a result, a tiny black hole (BH) is formed inside the NS. It grows by acquiring baryonic matter from the surrounding NS over a time period. Finally, the NS is transmuted into a BH with a mass similar to that of the NS, i.e., around $1$-$2\,M_\odot$. Observation of an old NS or the non-observation of a solar mass BH can constrain the DM-capture rate, in turn, the DM-neutron scattering cross-section $\sigma_{\chi n}$. 

In this work, we show that this scenario can drastically change if the DM particles are bosonic and form a Bose-Einstein condensate (BEC) inside the NS.  
As the BEC forms, the DM particles in the thermal core condenses into the ground state of the gravitational potential provided by the NS. The physical extension of the BEC state can be estimated using the Heisenberg uncertainty principle in the absence of large self-interaction. Therefore, the size of the DM core rapidly contracts from a thermal radius $r_\mathrm{th} \sim \mathcal{O}(10)\,\mathrm{cm}$ (at an age of $\sim 10^7$ years) to a much smaller radius, $r_\mathrm{BEC}\sim$ $10^{-5} ~\mathrm{cm}$ \cite{Bramante:2013hn}.
Therefore, we expect an enhancement in the total DM annihilation rate for a given s-wave cross section $\langle\sigma v\rangle$ due to the huge increase in the DM density in the dark core. As a result, we find that the surface luminosity of the NS may become detectable by the JWST even for significantly smaller values of $\expval{\sigma v}$, compared to Ref.\,\cite{Bramante:2013hn}. Fig.\,\ref{fig:illus} presents a schematic illustration of DM dynamics within a NS, highlighting the late-time temperature variation of NS with a BEC (III) and without a BEC core (II). 

We perform a detailed study on the implications of enhanced DM annihilation in the BEC state. First, we determine the time of BEC formation and the onset of enhanced annihilation by solving evolution equations of the NS temperature $\Tns$ and the DM number within the star. Afterward, we calculate the lower limit on $\sv$ for achieving the CA equilibrium and BH formation in a $10^{10}$ year-old NS. We find that the limit on $\sv$ is reduced by a factor of $\mathcal{O}(10^{10}-10^{15})$, compared to the scenario without BEC formation and enhanced annihilation. In most previous studies on DM annihilation-induced NS heating, the typical DM annihilation cross-section considered is $\sv \sim 10^{-26}{\,\rm cm^3 \,s^{-1}}$, which is motivated by the \emph{freeze-out} production of DM in the early Universe\,\cite{Lee:1977ua,Steigman:1984ac,Scherrer:1985zt,Jungman:1995df}. However, much smaller annihilation cross-section is plausible in \textit{feebly interacting} DM scenarios where DM is produced through the \emph{freeze-in} process in the early Universe\,\cite{Hall:2009bx,Yaguna:2011qn,Elahi:2014fsa,Bernal:2017kxu}. In these models, the annihilation cross-section can be several orders of magnitude smaller than the freeze-out case. Due to the enhancement of annihilation rate in the BEC state, \emph{the freeze-in DM models can be probed using JWST observation}. Here, for the first time in the \textit{DM capture scenario}, we constrain a freeze-in DM model with a scalar DM candidate and a scalar mediator using the projected sensitivity of JWST, together with the non-observation of transmuted black holes and the observed DM relic density.

This paper is organized as follows: the theory of DM capture and its subsequent dynamics leading to BEC formation have been discussed in Sec.\,\ref{sec:bec_dm}. Implications of enhanced annihilation on the NS temperature from BEC have been presented with numerical results in Sec.\,\ref{sec:sec3}. The model-independent constraints on DM annihilation cross-section and DM-neutron elastic scattering cross-section have been shown in Sec.\,\ref{sec:results}. In Sec.\,\ref{sec:self}, we briefly discuss how the present scenario can be modified in presence of non-zero DM self-interaction. In Sec.\,\ref{sec:model}, we provide a concrete realization of a freeze-in DM model where the BEC formation and enhanced annihilation is possible. Finally, we summarize our results and conclude in Sec.\,\ref{sec:discussion}.

%%%%%%%%%%%%%%%%%%%%%%%%%%%%%%%%%%%%%%%%%%%%%%%%%%
\section{Dark matter condensate in neutron star}     
\label{sec:bec_dm}
%\subsection{Enhanced dark matter annihilation from condensate}

If DM interacts with the nucleons inside a NS, then the DM particles from the surroundings will scatter and lose energy to get captured by the NS.
The DM capture rate of a NS of mass $\Mns$ and radius $\Rns$, residing in a DM local density of $\rho_\chi$ is given by\,\cite{Kouvaris:2007ay,Bell:2013xk},
\begin{align}
C_c&=\sqrt{6\pi}\left(\frac{\rho_\chi}{m_\chi \bar{v}_\chi}\right)\left(\frac{2G\Mns \Rns}{1-\frac{2G\Mns}{\Rns}}\right)f,
\label{eq:cap}
\end{align} 
where $f={\rm Min}[1,\sigma_{\chi n}/\sigma_g]$. Here, $\sigma_g$ is the geometric cross-section of the DM-neutron elastic scattering, and $\bar{v}_\chi$ is the DM average velocity in the DM halo. Since neutrons in the NS are highly degenerate, the capture rate is suppressed as limited phase space is available for the up-scattered neutrons. This effect is taken into account by parametrizing the geometric cross-section as follows\,\cite{Bell:2013xk}
\begin{align}
\sigma_g = \frac{\pi R^2_{NS}}{N_n} \left({\rm Min} \left[\frac{m_\chi}{0.2 {~\rm GeV}},1 \right]\right)^{-1}.
\label{eq:geom}
\end{align} 
For improved treatment of DM capture rate over a broad range of DM masses, including effects of relativistic energy transfer, Pauli blocking and multiple scatterings (particularly for DM mass above a PeV) see Refs.\,\cite{Bell:2020jou,Bramante:2017xlb,Dasgupta:2019juq}. Captured DM particles undergo further scattering inside the NS, and eventually thermalize with the environment to form a thermal dark core. The thermal radius $r_{\rm th}$ of the core is given by,
\begin{align}\label{eq:r_therm}
    r_{\rm th} = \sqrt{\frac{9\Tns}{4\pi G\rho_c m_\chi}} = 10\,{\rm cm}\left(\frac{\Tns}{10^4~\rm K }\right)^{1/2}\left(\frac{m_\chi}{100~\rm GeV}\right)^{-1/2},
\end{align}
where $\rho_c \approx 5\times 10^{38} {~\rm GeV\,cm^{-3}}$ is the NS core density. The dark core is stabilized by a balance between the thermal kinetic energy of DM particles and the gravitational potential energy due to baryons inside the NS\,\cite{Goldman:1989nd},
\begin{align}\label{eq:grav_energy}
    V(r) = \frac{4\pi G\rho_c}{3} m_\chi r^2 \,.
\end{align}
Here we assume a constant density $\rho_c$ at the core of the NS.

If DM particles are bosonic in nature, they can form a BEC if the temperature falls below a critical value $T_c$. Because the DM particles are trapped in the gravitational potential energy given by Eq.(\ref{eq:grav_energy}), we can treat them as a bosonic gas in a harmonic potential\,\cite{1999RvMP...71..463D,PhysRevA.35.4354}. The critical temperature for BEC formation in such a system is given by \cite{Jamison:2013yya}, 
\begin{align}\label{eq:Tc}
    T_c =\left(\frac{8\pi G \rho_c}{3}\right)^{1/2} \left( \frac{N_\chi}{\zeta(3)}\right)^{1/3},
\end{align}
where $N_\chi$ is the captured DM number inside the NS. Notably, this expression of $T_c$ is valid only for weakly interacting particles. This is true as long as DM particles are not self-gravitating. The critical number of DM particles for the onset of self-gravitation can be found by equating the DM density with that of the NS core, namely, $N_{\rm sg}=(4/3)\pi r^3_{\rm th}\rho_c/m_\chi$\,\cite{Acevedo:2020gro}. In the absence of additional self-interaction, for $N_\chi \ll N_{\rm sg}$, the inter-particle separation ($\sim N^{-1/3}_\chi$) remains larger than the typical length scale of self-gravitation ($~N^{-1/3}_{\rm sg}$), thereby the  assumption of weakly interacting gas is well justified. 

Moreover, when $N_\chi$ reaches $N_{\rm sg}$, the dark core is destabilized and begins to collapse under its own gravity, rather than forming a BEC state. Hence, the formation of a BEC state indeed takes place in the weakly interacting approximation in the present scenario. Comparing $N_{\rm sg}$ with the critical DM number required for the BEC formation, using Eq.(\ref{eq:Tc}), we find an upper limit on the DM mass,
\begin{align}
    m_\chi \lesssim 50 \,{\rm TeV} \left(\frac{\Tns}{10^4~\rm K}\right)^{-3/5}.
    \label{eq:bec}
\end{align}
Above this mass, the dark core collapses without forming a BEC. Therefore, our subsequent analyses have been performed for DM masses below 50 TeV. 

The physical size of the BEC, denoted by $r_{\rm BEC}$, formed by DM particles is much smaller than the thermal radius of the same number of DM particles. In a BEC, almost all particles occupy the ground state and the size of the BEC can be estimated using the Heisenberg uncertainty principle. It yields $r_{\rm BEC}=1/\sqrt{2m_\chi E_g}$ where $E_g$ is the energy of the ground state. Inside the NS, the ground state energy in the gravitational potential of Eq.(\ref{eq:grav_energy}) is $E_g= \frac{3}{2}\left(8\pi G\rho_c/3\right)^{1/2}$. Hence, the dark core radius is given by, 
\begin{align}\label{eq:r_bec}
   r_{\rm BEC}= \left(\frac{1}{24\pi G m^2_\chi \rho_c}\right)^{1/4} = 9.2\times 10^{-6}\,{\rm cm} \left(\frac{m_\chi}{100 ~\rm GeV}\right)^{-1/2}.
\end{align}
Therefore, it is evident from Eq.(\ref{eq:r_bec}) and Eq.(\ref{eq:r_therm}) that $r_{\rm BEC}\ll r_{\rm th}$ for the DM mass range we are considering. As the core of the NS cools below the critical temperature $T_c$ while DM particles are accumulated in the core, the thermal DM core makes a transition to a BEC with a much smaller size. As a result, the effective density of the DM particles goes up significantly. This, in turn, will enhance the annihilation rate of the DM particles inside the core. Here, we assume that the entire captured DM population reaches thermal equilibrium and subsequently condenses into the Bose–Einstein condensate state. Therefore, the total annihilation volume is determined only by $r_{\rm BEC}$, leading to the maximal enhancement in the DM annihilation rate. For details about the thermalization of captured DM particles see Refs.\cite{Bertoni:2013bsa,Garani:2018kkd,Bell:2023ysh}.

The assumption of a single-core BEC is formation from the initial thermal core is justified as the single-particle ground state in a harmonic trap potential is non-degenerate and the equilibrium condensate occupies the single lowest energy mode \cite{Pethick,PhysRevA.74.033612}. Examples of multiple-core BEC formation have been observed in laboratory for the double-well potential, and rotating Bose gases, due to their nearly degenerate ground states\,\cite{PhysRevA.74.033612}. However, we do not consider these possibilities in this work.

In passing, we also note that the timescale for the collapse of the dark core radius from its thermal value to the condensate is rather small compared to the other timescales (such as the change in the NS temperature), making it an almost instantaneous process. This time-scale can be estimated as $t_s = (r_{\rm th}-r_{\rm BEC})/v_s \approx r_{\rm th}/v_{\rm s} \sim 10^{-4}$ sec, where $v_{\rm s} \simeq \sqrt{\Tns/m_\chi}$ is the sound speed in the dark core. As long as the dark core is stabilized, the DM density becomes constant after the BEC formation, because the corresponding volume becomes independent of the NS temperature.

\section{Effect of enhanced annihilation on neutron star temperature}
\label{sec:sec3}
%\subsection{Neutron star core temperature evolution}
The formation of BEC takes place inside the NS when the core temperature falls below the critical temperature, i.e., $\Tns < T_c$. Initially, the evolution of $\Tns$ is controlled by the neutrino cooling (dominantly the modified-URCA process\,\cite{Shapiro:1983du}) and later by photon cooling. After the captured DM number grows to a sufficiently large value, they will start annihilating into SM particles which will act as a heating source for the NS. Therefore, the evolution of $\Tns$ is given by
\begin{align}
\frac{d\Tns}{dt} &= -\frac{\epsilon_\nu + \epsilon_\gamma - \epsilon_\chi}{c_V},
\label{eq:temp}
\end{align}
where $\epsilon_\nu$, $\epsilon_\gamma$ are standard emissivities coming from neutrino and photon cooling channels respectively. Here, $c_V$ is the total specific heat having a contribution from the dominant constituents (i.e. neutron, proton, electron) of the NS. These quantities depend on $\Tns$ as follows\,\cite{Kouvaris:2007ay,Shapiro:1983du,Page:2004fy}:
\begin{align}
\epsilon_\nu &= 3.3 \times 10^{-15} ~{\rm MeV^4~ year^{-1}} \left(\frac{\Tns}{10^{7} {~\rm K}}\right)^8 ,\nonumber\\
\epsilon_\gamma &= 1.7 \times 10^{-7} ~{\rm MeV^4~ year^{-1}} \left(\frac{\Tns}{10^{7} {~\rm K}}\right)^{2.2} ,\nonumber\\
c_V &= \frac{\Tns}{3}\sum_{i=n,p,e}p_{F,i}\sqrt{p^2_{F,i}+m^2_i}\,.
\end{align}
Here $p_{F,i}$ is the Fermi-momentum of the $i^{\rm th}$ particle species, which is dependent on its number density inside the star. In Eq.\eqref{eq:temp}, $\epsilon_\chi$ is the emissivity due to DM annihilation within radius $r$, given by
\begin{align}
\epsilon_\chi =  \frac{m_\chi C_a}{\frac{4}{3}\pi R^3_{\rm NS}} N^2_\chi,\qquad C_a=\frac{\expval{\sigma v}}{ (2\pi)^{3/2} r^3}.
\label{eq:demi}
\end{align} 
\begin{figure}[t]
    \centering
    \includegraphics[width=0.5\columnwidth]{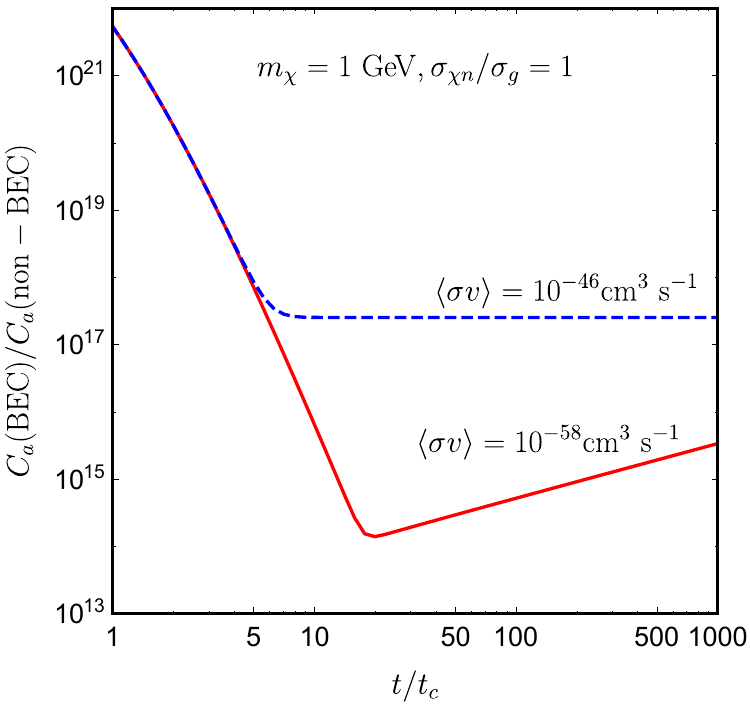}
    \caption{The enhancement factor defined in Eq.\eqref{eq:enhance} in DM annihilation rate shown as a function of scaled time $t/t_c$, where $t_c$ is the time of BEC formation, for two typical values of DM annihilation cross-section, $\expval{\sigma v}$ . The dependence on DM parameters in this factor comes through $\Tns$, determined by Eq.(\ref{eq:temp}) and Eq.(\ref{eq:num}). Here $\sigma_g$ is the geometric cross section  of the DM-neutron elastic scattering.}
    \label{fig:enhance}
\end{figure}
As mentioned before, when $N_\chi$ is sufficiently large, $\epsilon_\chi$ creates a new channel for NS heating.
The evolution of $N_\chi$ inside the NS is given by
\begin{align}
    \frac{dN_\chi}{dt} &= C_c-C_a N^2_\chi\,.
    \label{eq:num}
\end{align}
Initially, when the captured DM number $N_\chi$ is small, the 2nd term on the RHS of Eq.(\ref{eq:num}) is negligible, and the DM particles keep accumulating through capture. The annihilation rate eventually increases with $N_\chi$ and balances with the capture rate creating a capture-annihilation (CA) equilibrium.

This CA equilibrium can be reached faster if a BEC forms for bosonic DM as the annihilation rate increases by many orders of magnitude in this scenario for the same value of $\expval{\sigma v}$. The enhancement factor is given by
\begin{align}
\frac{C_a ({\rm BEC} )}{C_a ({\rm non-BEC})} = \left(\frac{r_{\rm th}}{r_{\rm BEC}}\right)^3 = 10^{18}\, \left(\frac{\Tns}{{10^4 \,\rm K}}\right)^{3/2} 
\label{eq:enhance}
\end{align}
The enhancement factor depends on the NS temperature through the temperature dependence of the thermal radius of the dark core only. The annihilation rate does not depend on temperature in the BEC state, unlike in the non-BEC state, where the annihilation rate increases with a decrease in $\Tns$. This stems from the fact that virtually all particles in the BEC state are in the ground state whose spatial extent is determined by the zero point energy. Here, we consider velocity-independent s-wave cross-section for which its thermal average, $\expval{\sigma v}$ remains unchanged, thereby the enhanced annihilation corresponds only to increased DM density in the core. Note that, in the condensate state, most of the particles have negligible velocities. Therefore, a $p$-wave cross-section, proportional to $v^2$, is likely to be strongly suppressed to produce a comparable enhancement.

Fig.\,\ref{fig:enhance} shows the enhancement factor for two different annihilation cross-sections. The large enhancement in the annihilation rate compared to the non-BEC case is clearly evident from the figure. In particular, the enhancement factor follows the evolution of $\Tns$ as apparent from Eq.\eqref{eq:enhance}. In the non-BEC case, the CA equilibrium is achieved with $\expval{\sigma v} = 10^{-46} {~\rm cm^3 ~s^{-1}}$; consequently $\Tns$ becomes constant at late times, as shown in Eq.\eqref{eq:tempf}. However, the CA equilibrium can not be achieved with $\expval{\sigma v} = 10^{-58} {~\rm cm^3 ~s^{-1}}$,  which results in a smaller $\Tns$ at later times compared to the former. Therefore, the enhancement factors differ approximately by a factor $\mathcal{O}(10^2)$ for these two cross-sections.       

As a result of the enhanced annihilation rate, the CA equilibrium is achieved with smaller values of $\expval{\sigma v}$, compared to the non-BEC case. In particular, we expect a substantial relaxation in the minimum value of the s-wave annihilation cross-section, i.e. around $10^{-54} {~\rm cm^3 ~s^{-1}}$, found earlier in Ref.\,\cite{Garani:2020wge}. Therefore, it implies that the \textit{feebly} interacting DM models with very small annihilation cross-section can be probed using the observation of old NSs. After reaching the CA equilibrium, $\Tns$ gets frozen at a constant value $T^{eq}_{\rm NS}$ as follows\,\cite{Kouvaris:2007ay}
\begin{align}
T^{eq}_{\rm NS} \approx 1.98\times 10^{5} {~\rm K}~\left(\frac{m_\chi}{{10~\rm GeV}} \times\frac{C_c}{8\times 10^{35} {~\rm year^{-1}}}\right)^{0.45},
\label{eq:tempf}
\end{align}
which is related to the NS surface temperature $T_{s}$ due to the high conductivity between the core and the crust of the star. The relation is given by the following\,\cite{Gudman1,Gudman2,Kouvaris:2007ay},
\begin{align}
 T_s = 9.1 \times 10^{5} ~{\rm K}~ \left(\frac{\Mns}{M_\odot}\right)^{1/4}\left(\frac{10~ \rm km}{\Rns}\right)^{1/2} \left(\frac{\Tns}{9.5\times 10^{7}~{\rm K}}\right)^{0.55},
\end{align}
which is valid for $\Tns \gtrsim 3000$ K, below which one can set $T_s\simeq\Tns$ \cite{Bramante:2023djs}.

Another interesting phenomenon in slowly annihilating DM models is dark core collapse in the BEC state, leading to the formation of a solar mass black hole after devouring the host star\,\cite{Bramante:2013hn,Dutta:2024vzw}. The existence of old NSs puts a lower limit on the DM annihilation rate for a given DM-neutron scattering cross-section, which now will be modified due to enhancement in DM annihilation from a BEC.  The gravitational collapse of the DM core will happen when $N_\chi \gtrsim N_{\rm ch}$, where $N_{\rm ch} = 16 M^2_P/m^2_\chi$ is the \textit{Chandrasekhar limit} \cite{Bramante:2013hn}.
This is the critical number of particles for core collapse in the BEC state, which is typically smaller than that in the non-BEC case, $N_{\rm sg}$. This is due to the fact that in the non-BEC case, thermal pressure maintains the gravitational stability of the core. Whereas in a BEC, the gravitational stability is supported by the quantum pressure only, which can be overcome with less number of particles. Therefore, the effects of enhanced annihilation in the NS temperature profile will be observable for $ N_\chi < N_{\rm ch}$ at temperatures $\Tns < T_c$. Since the dark core radius becomes constant for $\Tns<T_c$, the evolution of the DM number becomes independent of $\Tns$ and  we find its analytical solution as follows,  
\begin{align}
    N_\chi (t) = \sqrt{\frac{C_c}{C_a}} \frac{A \exp(2\sqrt{ C_a C_c} (t-t_c))-1}{A \exp(2\sqrt{ C_a C_c} (t-t_c))+1},~~\quad
    A = \frac{1+\sqrt{\dfrac{C_a}{C_c}}N_\chi(t_c)}{1-\sqrt{\dfrac{C_a}{C_c}}N_\chi(t_c)},
    \label{eq:EAnum}
\end{align}    
where $t_c$ is the time of BEC formation, calculated numerically by solving Eq.\eqref{eq:temp} and Eq.\eqref{eq:num} simultaneously. Subsequently, the late-time behavior of $\Tns$ is computed by solving Eq.\eqref{eq:temp} using the above expression. 

\section{Implications for dark matter parameters}
\label{sec:results}
In this section, we systematically study the effect of enhanced annihilation in the BEC state in determining the CA equilibrium, black hole formation, and late-time heating effect. We do so by solving the NS temperature $\Tns$  and the captured DM number $N_\chi$ equations, i.e., Eq.\eqref{eq:temp}, \eqref{eq:num}, and \eqref{eq:EAnum}, for a NS lifetime of $10^{10}$ years. We take initial conditions as, $\Tns=10^{10}$ K and $N^i_{\chi}=0$ at $t=100$ years when the NS temperature becomes only a function of time. We first determine the time of BEC formation for a set of DM parameters and, if BEC forms, we assume instantaneous onset of enhanced annihilation. Our results are shown for typical values of NS parameters: $\Mns =1.4 M_\odot$, $\Rns=10.6$\,km. We choose ambient DM density $\rho_\chi=0.4\,{\rm GeV cm^{-3}}$ and the average DM velocity, $\bar{v}_\chi=220 ~{\rm km~ s^{-1}}$\,\cite{deSalas:2020hbh}.   
We have used the observation of the pulsar PSR J2124-3358 with a spin-down age of $7.2\times 10^9$ years  as a benchmark\,\cite{Mignani:2003nw,Garani:2018kkd}.
\begin{figure}[t]
%\hspace{-1.2 cm}
    \includegraphics[width=0.49\columnwidth]{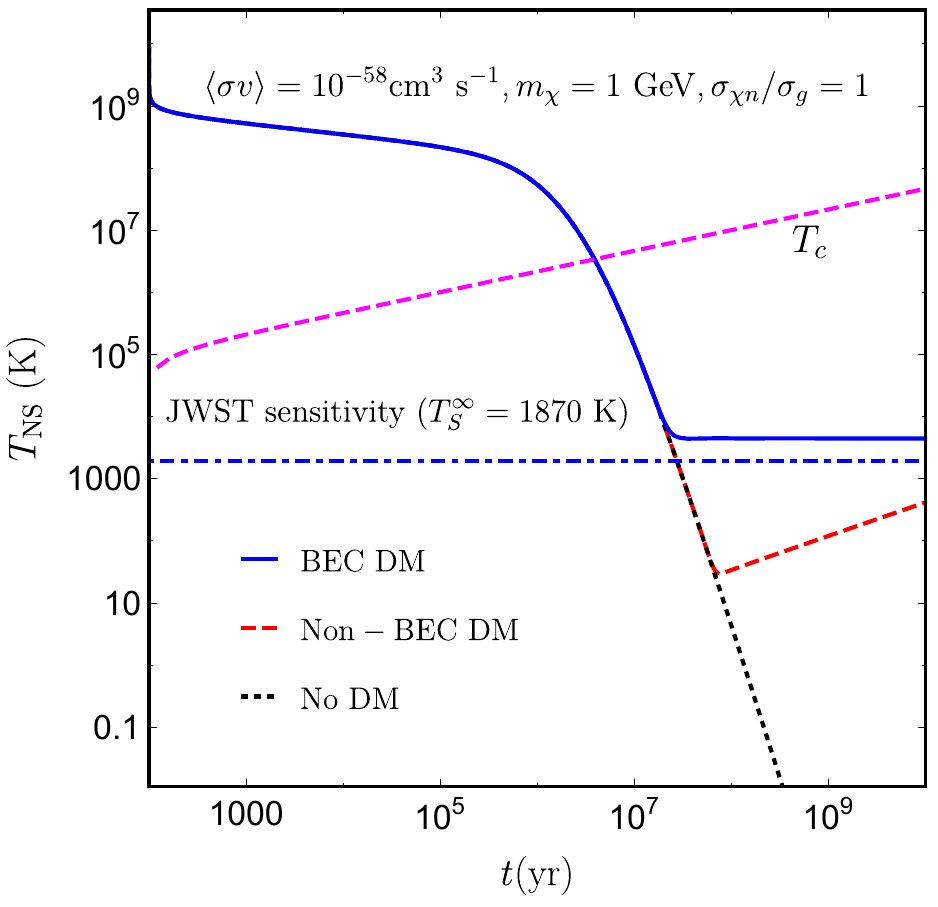}
    \includegraphics[width=0.48\columnwidth]{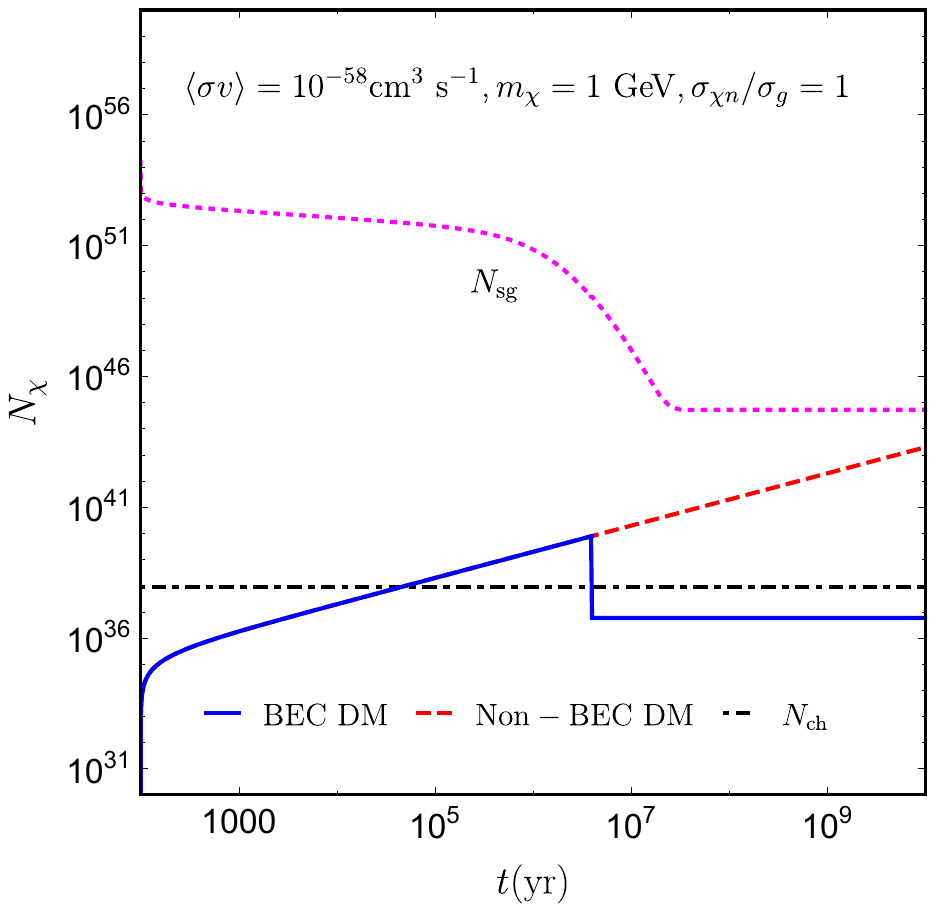}
    \caption{Evolution of the NS temperature (left panel) and the captured DM number (right panel) with time showing the effect of enhanced annihilation of DM in the condensate state. In the BEC case (solid blue), the capture-annihilation equilibrium is established after the condensate formation. Whereas, for the same annihilation cross-section, the equilibrium is not achieved in the non-BEC scenario (dashed red). As an observational consequence, the BEC scenario NS surface temperature remains above the JWST sensitivity, while the non-BEC case remains unobservable. In the left panel, the dashed magenta line shows the evolution of the critical temperature $T_c$. In addition, NS stability is also ensured since the DM number remains smaller than the critical number for dark core collapse for $10^{10}$ years of old NS in DM environment with $\rho_\chi = 0.4~{\rm GeV cm^{-3}}$. In the right panel, the dotted magenta line shows the critical number $N_{\rm sg}$ of DM particles needed for gravitational collapse in the non-BEC case, and the gray, dot-dashed line shows the same number for the BEC scenario.}
    \label{fig:evol}
\end{figure}

In Fig.\,\ref{fig:evol}, we demonstrate the new effects due to enhanced annihilation of a GeV mass DM with $\sigma_{\chi n}/\sigma_g = 1$ and $\expval{\sigma v} = 10^{-58} ~\rm{cm^3 ~s^{-1}}$ in three aspects of standard DM capture scenarios in NS, namely - (1) CA equilibrium, (2) BH formation, and (3) late-time heating. To note, for $\sigma_{\chi n}=\sigma_g$, the entire DM population reaches condensation, giving rise to maximum effect of enhanced annihilation in all three categories.
In the left panel, we show the evolution of $\Tns$ with time for multiple scenarios. The evolution of the critical temperature $T_c$ for BEC formation as a dashed magenta line. In the case where a BEC forms (blue, solid), we note that $\Tns$ drops below $T_c$ at around $t\simeq 5\times10^6$\,years. As a result, enhanced annihilation starts affecting the evolution of $\Tns$, evident from the non-BEC case, shown by the red dashed line.

The captured DM number $N_\chi$ evolution is shown by blue solid line in the right panel of Fig.\,\ref{fig:evol}. Due to the enhanced annihilation at the onset of the BEC formation, $N_\chi$ drops by $\mathcal{O}(10^3)$ at around $t\simeq 5\times10^6$\,years. Afterward, the CA equilibrium is established and captured DM number stabilizes. Note that with the chosen values of the DM parameters, the CA equilibrium would not be achieved if there was no enhanced annihilation. The evolution of DM number without enhanced annihilation is shown as a red dashed line. It is only because $N_\chi$ falls below the Chandrasekhar limit $N_{\rm ch}$, the dark core stabilizes without collapsing into a BH even for such small $\expval{\sigma v}$. We have also shown the critical number of self-gravitation, $N_{\rm sg}$ (shown by the magenta dotted line), relevant for non-BEC state of DM, which is always greater than the DM number throughout the evolution, thereby ensuring the stability of the dark core. 

The enhanced DM annihilation will affect the observability of a NS as well. As the CA equilibrium is achieved, the late-time heating of NS becomes large enough to raise the NS surface temperature above the sensitivity of the JWST. This is demonstrated in the left panel of Fig.\,\ref{fig:evol}. The NS temperature stabilizes above the lowest temperature $\Tns=1870\,$K that JWST can observe, shown as a blue dot-dashed line. In fact, we find the NS temperature to be $\sim\mathcal{O}(10)$ times higher at $t=10^{10}$ years than the non-BEC case, which is shown as the red-dashed line.

Therefore, the DM BEC formation ensures that NS surface is `\textit{bright}' enough to be observed with a telescope which would not be possible otherwise. Moreover, this is possible with a much smaller DM annihilation and DM-neutron scattering cross-sections, that are not ruled out by indirect and direct detection experiments, respectively. In fact, for \mbox{$m_\chi = 1 $\,GeV,} $\sigma_{\chi n}$ can be increased to the geometric cross section $\sigma_g$, still allowed by all experiments, for which the capture rate will be maximum and the NS will be even brighter. This opens up the DM parameter space.
\begin{figure}[t!]
    \centering
    \includegraphics[scale=0.75]{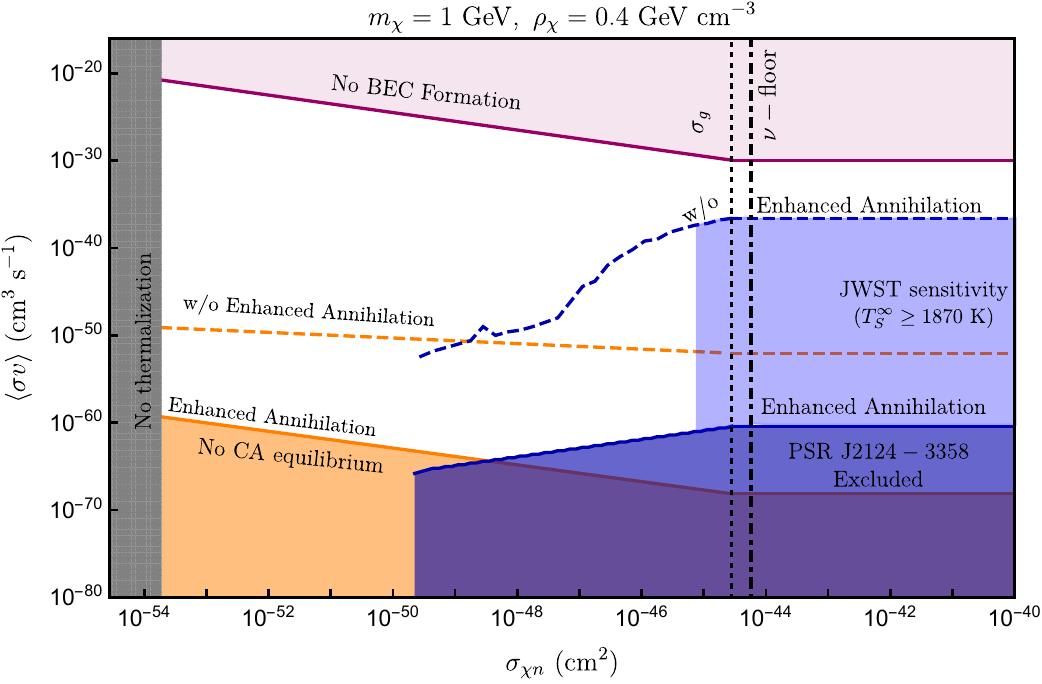}
    \caption{Changes in the $\expval{\sigma v}-\sigma_{\chi n}$ parameter space due to the enhanced annihilation from DM BEC formation. The blue solid line shows the $\expval{\sigma v}$ below which the DM core collapses into a BH when the enhanced annihilation is taken into account. This region is excluded by the observation of $7.2\times10^9$\,yrs old pulsar PSR J2124-3358. The blue dashed line shows the same cross-section without enhanced annihilation. The light blue-shaded region shows the new part of the parameter space which now can be probed by JWST as the NS does not collapse into a BH. The minimum $\expval{\sigma v}$ needed to establish CA equilibrium is shown as the solid orange line in the BEC case, which decreases by a factor of $10^{10}-10^{15}$ from the non-BEC case which is shown as the dashed orange line. BEC formation is not possible in the region above the purple line, and DM thermalization is not possible in the gray-shaded region. 
    }
    \label{fig:fig1}
\end{figure}

We show all of these changes between the \emph{with and without enhanced annihilation  cases} in Fig.\,\ref{fig:fig1} in the $\expval{\sigma v}-\sigma_{\chi n}$ plane  for 1\,GeV mass DM as an example. For values of $\expval{\sigma v}$ above the solid purple line, DM particles do not form a condensate as the critical number for condensation is never achieved in $10^{10}$ years. Clearly, this upper bound on $\expval{\sigma v}$ for BEC formation weakens for smaller values of $\sigma_{\chi n}$ as the capture rate decreases. 

We also update the region of the parameter space excluded by gravitational collapse criterion which is shown as a solid blue line in Fig.\,\ref{fig:fig1}. For a $\expval{\sigma v}$ below this line, the dark matter core will collapse to form a solar mass BH. This region is excluded by the observation of the pulsar PSR J2124-3358. As can be seen from Fig.\,\ref{fig:fig1}, this stability condition on $\expval{\sigma v}$ is relaxed by approximately $\sim15$\,-\,25 orders of magnitude due to the enhanced annihilation within the parameter space shown, relative to the case without enhanced annihilation, shown in dashed blue line. 
For higher values of $\sigma_{\chi n}$, BH forms earlier at higher NS temperature, consequently the annihilation rate within the thermal radius is suppressed, thereby the enhancement factor (see Eq.\eqref{eq:enhance}) becomes pronounced. Therefore, the change in the limit of $\expval{\sigma v}$ is maximum at the geometric cross-section, for which BH forms at the earliest.  

The lower limit on $\expval{\sigma v}$ required to achieve the capture-annihilation equilibrium within the lifetime of NS is shown for the with and without enhanced annihilation cases in Fig.\,\ref{fig:fig1} by the orange solid and dashed lines respectively. It is apparent that the CA equilibrium can be established with a $\expval{\sigma v}$ smaller by 10\,-15 orders of magnitude due to the enhanced annihilation from the BEC. This has significant observational implication for NSs as more parameter space opens up in $\expval{\sigma v}$ which will make old NSs observable due to additional heating from DM annihilation. 

\begin{figure}[t]
\centering
\hspace{-1.5 cm}
\includegraphics[scale=0.466]{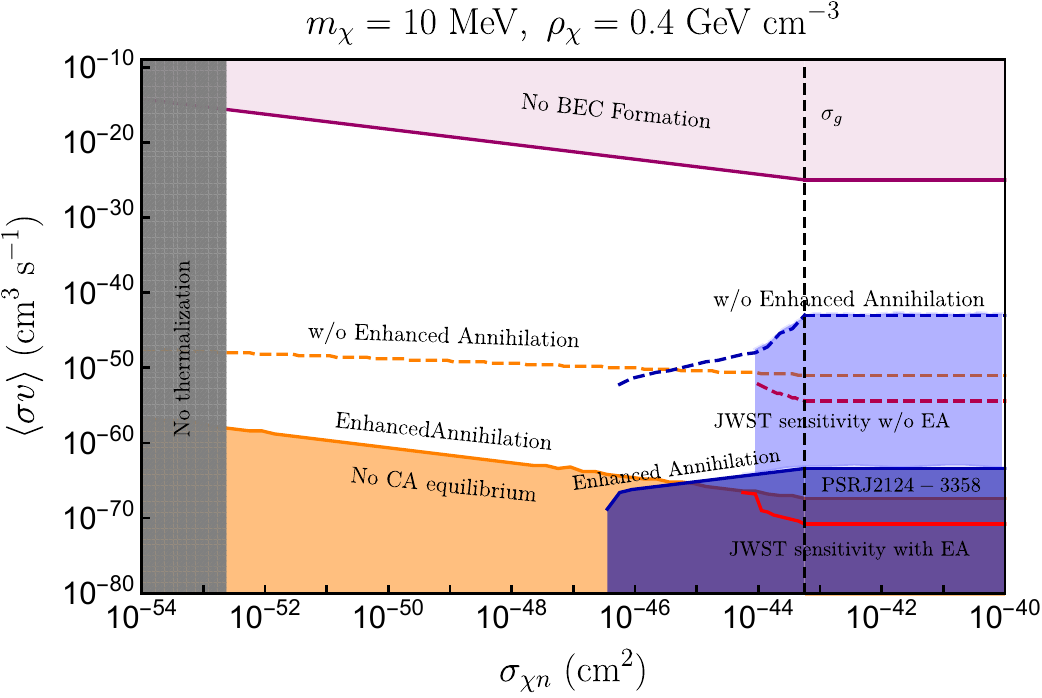}
\includegraphics[scale=0.455]{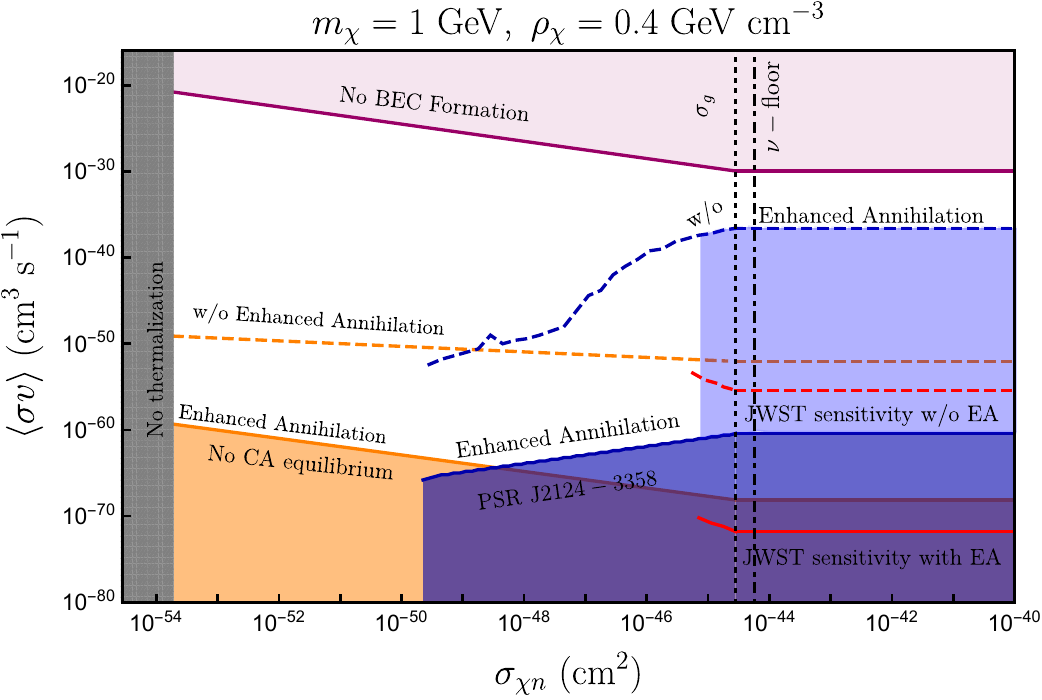} \\[0.5cm]
\includegraphics[scale=0.5]{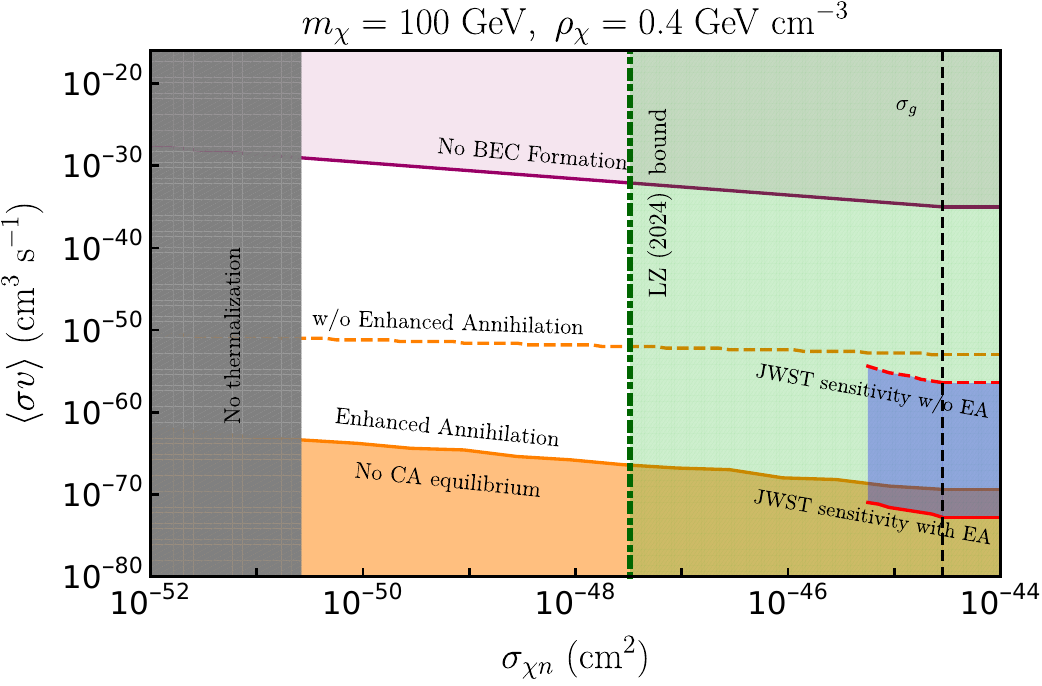}
 \caption{ Effects of enhanced annihilation (EA) due to DM BEC formation for $m_\chi= 10$ MeV (\emph{upper left}) $m_\chi = 1$ GeV (\emph{upper right}), and $m_\chi=100$ GeV (\emph{bottom}) similar to Fig.\,\ref{fig:fig1}. JWST sensitivity is shown by red solid and dashed lines if the NS did not collapse into a BH. For $m_\chi =100$ GeV, black hole does not form due to dominant Hawking radiation, thereby probing smaller $\expval{\sigma v}$. Direct  detection bound on $\sigma_{\chi n}$ becomes relevant for 100 GeV DM, as green shaded region is excluded from LZ experiment \cite{LZ:2024zvo}.  } 
    \label{fig:fig3}
\end{figure}

To explain this point more clearly, we have shaded the region in Fig.\,\ref{fig:fig1} in light blue in which the surface temperature of a $10^{10}$\,years old NS is within the observable limit of JWST due to enhanced annihilation. Without enhanced annihilation BH would have formed in this region. As BH is now formed with smaller $\expval{\sigma v}$, the NS remains stabilized with a core made of BEC DM. In addition, with the revised lower limit on $\expval{\sigma v}$ for the CA equilibrium, the measurement of $T_s$ can be possible with JWST, which is previously ruled out due to transmutation of the NS into a BH. The JWST sensitivity to $\expval{\sigma v}$, independent of NS transmutation, has been shown in Fig.\,\ref{fig:fig3}. We also note that the NS surface temperature falls below the JWST sensitivity for $\sigma_{\chi n} \lesssim 7\times 10^{-46}~{\rm cm^2}$, due to insufficient DM capture throughout the NS lifetime. However, we can probe $\sigma_{\chi n}$ below the \emph{neutrino floor} (also known as \textit{neutrino fog}) with observation of the surface temperature of a nearby old NS using infra-red capabilities of JWST. For $m_\chi=1$ GeV, the neutrino floor corresponding to a Xenon target is at $\sigma_{\chi n}\simeq 5.7\times 10^{-45}{~\rm cm^2}$ \cite{OHare:2021utq}, shown by the black dot-dashed line in Fig.\ref{fig:fig1}.   

The gray-shaded regions in Fig.\,\ref{fig:fig1} and also in Fig.\,\ref{fig:fig3} indicate $\sigma_{\chi n}$ values for which thermalization can not be achieved before the BEC formation. For a given set of DM parameters, the thermalization timescale should be smaller than the timescale related to the formation of BEC. Therefore, the minimum required value of  $\sigma_{\chi n}$ for DM thermalization has been estimated following Refs.\,\cite{Bramante:2013hn,McDermott:2011jp,Dutta:2024vzw}, comparing $\Tns$ at $t=10^8$ years with the critical temperature for the BEC transition, as shown in Eq.(\ref{eq:Tc}). In particular, for a 1 GeV DM, the minimum value of $\sigma_{\chi n}$ is obtained as, $\sigma_{\chi n}\simeq 2\times 10^{-54} ~{\rm cm^2}$.

\begin{figure}[t]
\centering
\hspace{-1.5 cm}
\includegraphics[scale=0.45]{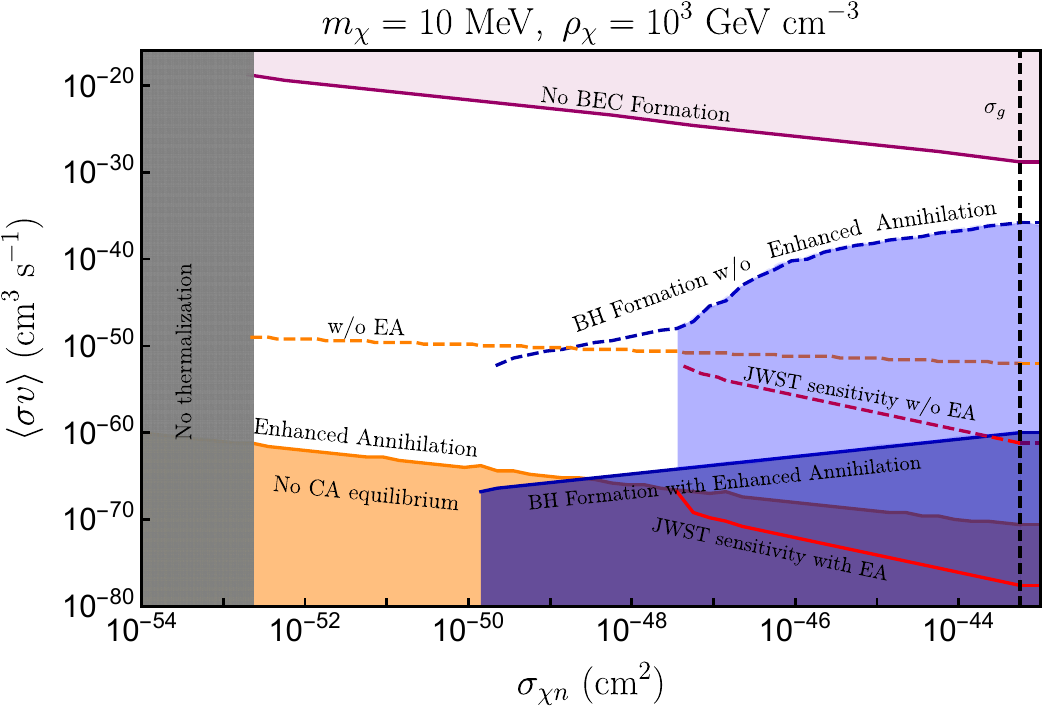}
\includegraphics[scale=0.46]{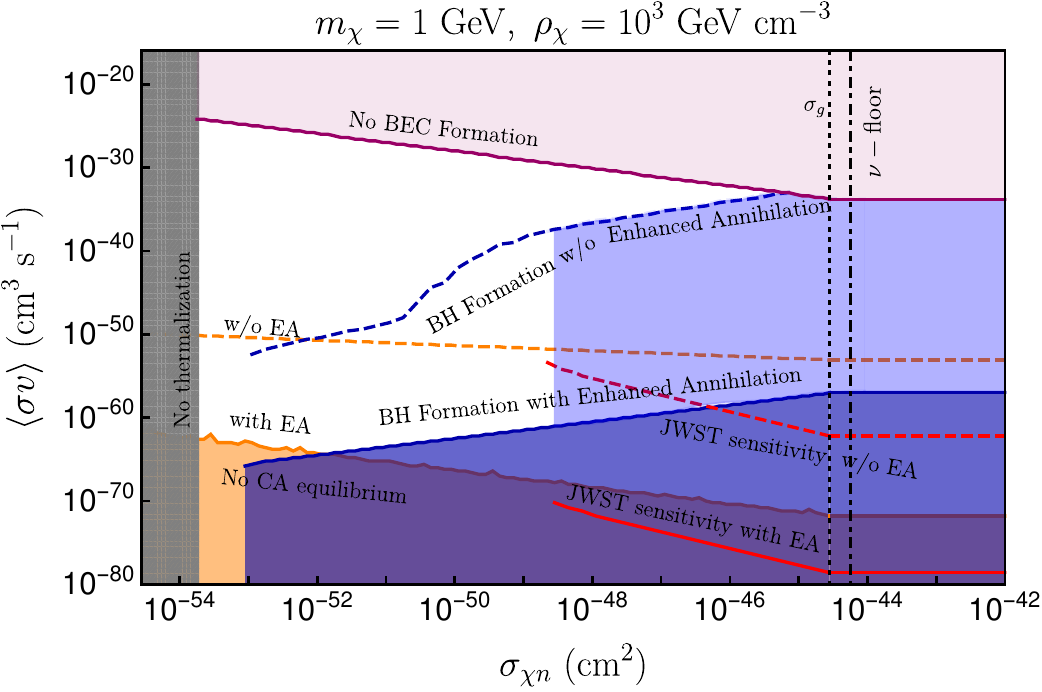} \\[0.5cm]
\includegraphics[scale=0.5]{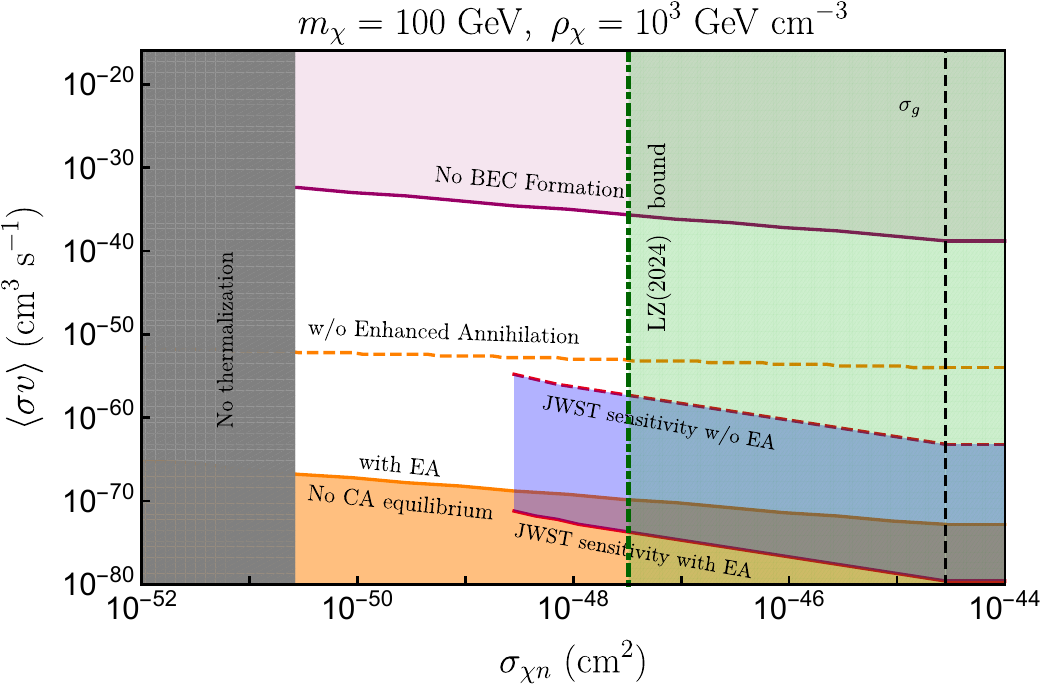}
 \caption{ Same figures as in Fig.\ref{fig:fig3} but for higher DM density $\rho_\chi=10^3\,{\rm GeV\,cm^{-3}}$, which allows us to probe even smaller values of $\sigma_{\chi n}$. For all limits, we have assumed an old NS with spin-down age of $10^{10}$ years.}  
\label{fig:1000}
\end{figure}
In the upper right panel of Fig.\,\ref{fig:fig3}, we present a detailed version of Fig.\,\ref{fig:fig1} to show how the JWST sensitivity to $\expval{\sigma v}$ improves with enhanced annihilation in the BEC state. In particular, with enhanced annihilation the JWST sensitivity goes down to $\expval{\sigma v}$ as small as $10^{-72} {~\rm cm^3~ s^{-1}}$ (see red solid line), which is approximately 15 orders of magnitude improvement from the without enhanced annihilation scenario, shown by the red dashed line. However for $m_\chi=10\,$MeV and 1\,GeV, the NS transmutes to a BH, thereby the detection of the surface temperature is possible only with higher DM cross-sections as depicted by the light blue shaded regions. For $m_\chi=10\,$MeV, shown in the upper left panel, the features of enhanced annihilation qualitatively remains the same except some changes in the limits on $\expval{\sigma v}$ due to smaller $m_\chi$. This is because the DM capture rate is enhanced as the number of DM particles in the vicinity of the NS scales as $1/m_\chi$. In addition, the geometric cross-section of the DM-neutron interaction increases for $m_\chi <0.2$\,GeV, requiring higher $\sigma_{\chi n} $ for the maximum  capture rate. Consequently, the DM capture rate increases approximately by a factor of 5 for $m_\chi=10\,$MeV compared to 1\,GeV for a given $\sigma_{\chi n}$. Thus, the upper limit on the DM annihilation for BEC formation becomes weaker in case of 10\,MeV mass, whereas it gets stronger for $m_\chi=100\,$GeV, since the capture rate is suppressed by a factor of 100. Furthermore, for $m_\chi=100\,$GeV, the host NS does not transmute into a BH since the seed BH from the dark core collapse would evaporate away within the lifetime of the star. In fact, for $m_\chi \gtrsim 10$ GeV, BH does not form out of a BEC state \cite{Kouvaris:2011fi,Garani:2018kkd}. We have also argued that DM particles do not become self-gravitating for $m_\chi \lesssim 50 $ TeV, as shown in  Eq.(\ref{eq:bec}). For $m_\chi=100$ GeV, the minimum $\expval{\sigma v}$ for the CA equilibrium is around $10^{-64}{~\rm cm^3~s^{-1}}$ with enhanced annihilation, which is smaller than other two masses.  Therefore, we expect to probe smaller $\expval{\sigma v}$ with JWST observation. However, the direct detection bound from  LZ experiment excludes $\sigma_{\chi n} \gtrsim 3.26 \times 10^{-48} {~\rm cm^2}$ for 100 GeV DM, in contrast to former case, where the direct detection bound is rather weak\,\cite{LZ:2024zvo}. As a consequence, the region that could be probed by JWST is already ruled out for $\rho_\chi=0.4 \,\rm GeV\,cm^{-3} $. For denser DM environment, the DM parameter space becomes accessible for $m_\chi=100$ GeV, as depicted in Fig.\,\ref{fig:1000}.

We also show the results for ambient DM density $\rho_\chi=10^3\,{\rm GeV cm^{-3}}$ in Fig.\,\ref{fig:1000}. In presence of DM spikes near the galactic center, such values of DM density can be realized in the outer region of the spike, i.e. $10^{-2}$ pc from the Galactic Center\,\cite{PhysRevLett.83.1719,Fields:2014pia}. Additionally, similar DM density can be found in globular clusters, like M4, while DM velocity is found to be around $20 {~\rm km~s^{-1}}$ assuming the King model \cite{King} of stars within the cluster \cite{Bertone:2007ae}. As is evident from Eq.\eqref{eq:cap}, the large $\rho_\chi$ can compensate for smaller $\sigma_{\chi n}$ in the DM capture rate, JWST can probe  much smaller cross section $\sigma_{\chi n}$ which can be seen in Fig.\,\ref{fig:1000}.

To summarize, we find that, due to the enhanced annihilation, the minimum required DM annihilation cross-section to reach the detection threshold of JWST becomes much smaller compared to the canonical WIMP scenario \cite{Lee:1977ua,Steigman:1984ac,Scherrer:1985zt}. In Sec.\ref{sec:model}, we discuss a concrete DM model where such small cross-sections can be achieved.

\section{Beyond the ideal condensate approximation}
\label{sec:self}
Till now, we have discussed the effect of BEC formation out of DM particles in the absence of self interaction. In presence of sizable DM self-interaction, the BEC formation and its subsequent dynamics can be different. At first, we must check whether the self-interaction can substantially change the critical temperature of the BEC transition. The effect of self-interaction in deciding the critical temperature of the transition can  be neglected if $n_\chi a^3 \ll 1$, where $n_\chi$ is the DM number density inside the thermal core and $a$ is the s-wave self-scattering length. The shift in the critical temperature for dilute interacting Bose gas is given by, $\frac{T_c-T^0_c}{T_c} \simeq 2.9 n_\chi a^{1/3}$, where $T^0_c$ is the critical temperature for the Bose gas without self-interaction \cite{PhysRevLett.83.1703, PhysRevB.80.014502}. Using the Bullet cluster bound on the DM self-scattering cross-section, i.e. $\sigma_{\chi\chi}/m_\chi \leq 1 \rm cm^2/gm$ \cite{2008ApJ...679.1173R,Robertson:2016xjh}, we can estimate the scattering length as,
\begin{align}
    a= \left(\frac{\sigma_{\chi\chi}}{4\pi }\right)^{1/2}= 3.7\times 10^{-13} ~{\rm cm} \left(\frac{\sigma_{\chi \chi}/m_\chi}{1 \rm cm^2/gm}\right)^{1/2}\left(\frac{m_\chi}{\rm GeV}
    \right)^{1/2}.
\end{align}
Before the BEC transition, DM particles are confined within the thermal core, for which the DM density, $n_\chi=3N_\chi/(4\pi r^3_{\rm th})$. Hence, the self-scattering effect is negligible as long as
\begin{align}
N_\chi \ll 8.3\times 10^{46} \left(\frac{\sigma_{\chi \chi}/m_\chi}{1 \rm cm^2/gm}\right)^{-3/2} \left(\frac{\Tns}{10^6 ~\rm K}\right)^{3/2} \left(\frac{m_\chi}{\rm GeV}\right)^{-3},
\end{align}
even with largest possible DM self-interaction strength. As shown in Fig.\ref{fig:evol}, the BEC transition occurs around $T_{\rm NS} =10^6$ K and $N_\chi\simeq 10^{40}$ for $m_\chi=1$ GeV, thus $T_c$ is determined only by the harmonic potential. Furthermore, with smaller DM masses, the required DM number is even higher to substantially alter $T_c$.  Therefore, for a thermal DM inside the NS, we safely neglect the effect of self-interaction before the BEC formation.  
\begin{figure}[t]
\hspace{-1.5 cm}
    \includegraphics[scale=0.5]{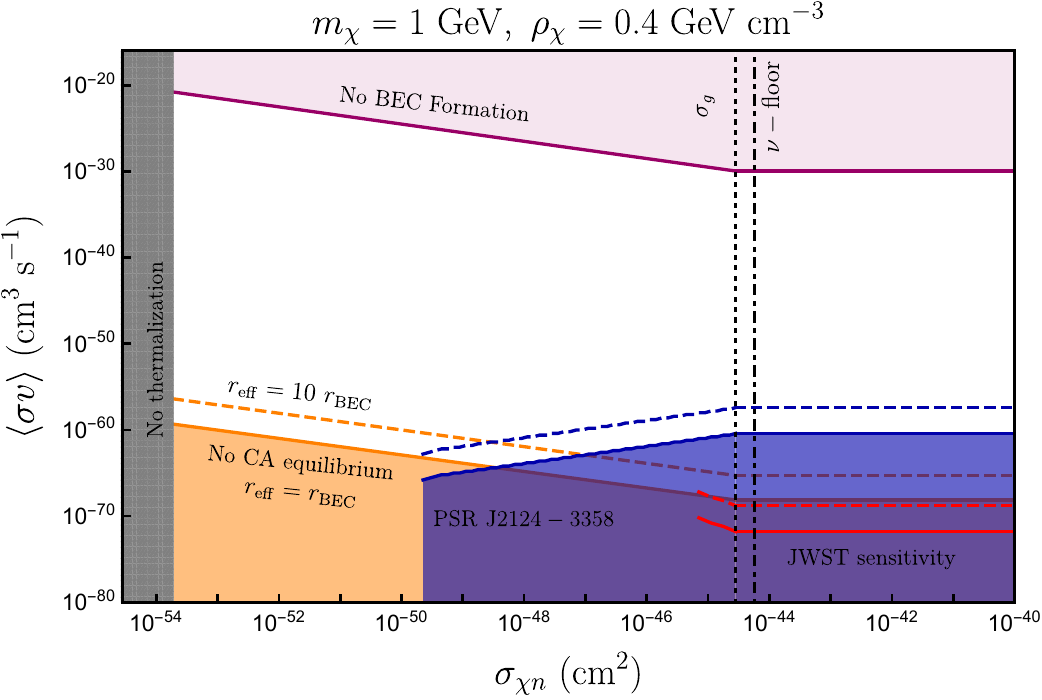}
     \includegraphics[scale=0.5]{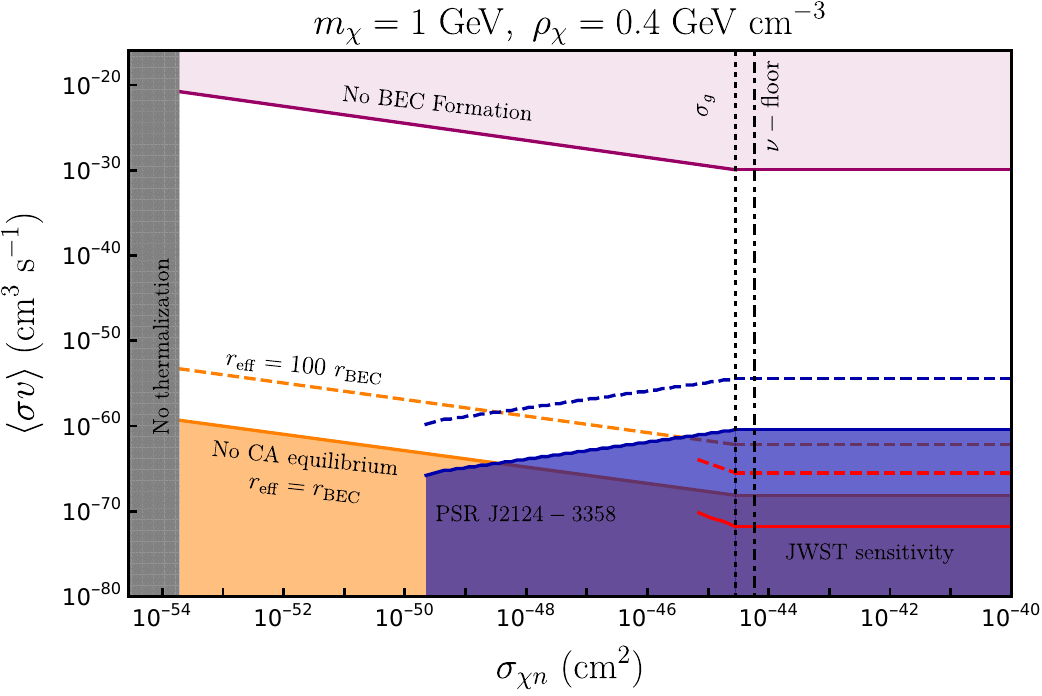}
    \caption{\textit{Comparison between ideal and non-ideal BEC cases.} In presence of repulsive self-interaction, the radius of the DM BEC core can be enhanced to $r_{\rm eff} > r_{\rm BEC}$. Here, we show the modification in bounds on $\expval{\sigma v}$ for two benchmark two cases with $r_{\rm eff} =10 r_{\rm BEC}$ (left panel) and $r_{\rm eff} = 100 r_{BEC}$ (right panel) in determining the CA equilibrium, stability of the NS and the JWST sensitivity of the NS surface temperature. Modified bounds are shown in dashed lines, while solid lines signify the $r_{\rm eff} =r_{\rm BEC}$ case. }
    \label{fig:self}
\end{figure}

The effect of self-interaction can be important after the BEC transition as the DM density grows as a result of volume contraction of the dark core. For repulsive self-interaction, the radius of the dark core increases than in the ideal BEC case, depending on the ambient DM number inside the NS. For an attractive self-interaction, the stability of the condensate can be perturbed once the DM number crosses a certain critical value \cite{Pethick}.  However, in the present case, the situation is not so straightforward, primarily due to the presence of DM capture and annihilation. There can be interplay of self-interaction strength with interaction energy related to DM acquisition and dissipation within the condensate, which potentially can modify the radius of the condensate or its stability from usual treatment only with self-interaction \cite{Ueda_1999,Donley_2001, PhysRevLett.81.933}. To treat the condensate dynamics consistently, we need to solve non-linear Gross–Pitaevskii (GP) equation with terms contributing from DM capture, annihilation and self-interaction, which requires a separate dedicated study. In particular, for attractive self-interaction, the stability criterion is expected to be modified due to the dissipative term in the GP equation, stemming from DM annihilation.

In Fig.\ref{fig:self}, we show the effects of possible change in the dark core in presence of repulsive self-interaction, by comparing the effective dark core radius, $r_{\rm eff} =10 ~r_{\rm BEC}, 100 ~r_{\rm BEC}$ with the ideal BEC case with $r_{\rm eff} = r_{\rm BEC}$. The Dashed lines show the modified bounds on $\expval{\sigma v}$ for the non-ideal BEC case, whereas the solid lines correspond to the ideal BEC as in the previous section. Since the enhanced annihilation rate is reduced by factors of $10^3$ and $10^6$ for two cases shown in the figure, the capture-annihilation equilibrium and the BH formation happen for a larger $\expval{\sigma v}$ by the same amount.
\section{A freeze-in model : Scalar dark matter with scalar portal}
\label{sec:model}
In this section, we introduce a concrete dark sector model where the enhanced annihilation from a dark matter condensate inside a NS can be realized. We consider a real scalar singlet, $\chi$ and  a scalar mediator, $\phi$ which has a trilinear interaction with $\chi$ and Yukawa interaction with SM quarks $f$ as shown below \cite{Blennow:2013jba},
\begin{align}
    -\mathcal{L}_{\rm int} \supset \frac{\mu}{2}  \phi \chi^2 + \lambda_f \phi \bar{f}f \,.
\end{align}
Here, we take the same Yukawa coupling $\lambda_f$ for all generations of quarks for simplicity. Note that both $\chi$ and $\phi$ can have interactions with SM Higgs through terms like- $\phi H^\dagger H$, $\phi^2 H^\dagger H$ and $\chi^2 H^\dagger H$. However, we assume these couplings to be small. The DM $\chi$ is stabilized by a discrete $\mathcal{Z}_2$-symmetry. As a primary representation of our model-independent results, we show that sufficient DM-neutron elastic scattering cross-section, i.e $\sigma_{\chi n} \sim \mathcal{O}(10^{-35}-10^{-55}){~\rm cm^2}$ and  feeble DM annihilation cross-section, i.e., $\langle\sigma v\rangle \sim\mathcal{O}(10^{-40}-10^{-80}){~\rm cm^3~ s^{-1}}$ is achievable within this realistic setup with suitable choice of the model parameters. 

\subsection{Dark matter-nucleon scattering cross-section}
First, we calculate the elastic scattering cross-section between DM and neutron, which is facilitated by a t-channel process mediated by $\phi$. The scattering amplitude of $\chi(p_1)+ n(p_2) \rightarrow \chi(p_3)+n(p_4)$ is given by,
\begin{align}
    i\mathcal{M} = \frac{-i\mu \lambda_f}{t-m^2_\phi} F_N \bar{u}_n (p_2) u_n(p_4),
    \label{eq:amp}
\end{align}
where $F_N$ is form factor accounting for transition from a quark-level operator ($\bar{q}q$) to a nucleon-level operator ($\bar{n}n$). Both light quarks and heavy quarks (via QCD trace anomaly) contribute to $F_N$ as follows \cite{Hisano:2015bma,Ellis:2008hf},
\begin{align}
    F_N=
    \sum_q\langle n(k)|\bar{q} q| n(k^\prime)\rangle = m_n\left[ \sum_{u,d,s} \frac{f^n_{T,q}}{m_q} +\frac{2}{27} \left(1-\sum_{u,d,s}\frac{f^n_{T,q}}{m_q}\right)\right] \approx 15 ~\frac{m_n}{ {\rm GeV}},
\end{align}
where $f^n_{T,u}=0.013, f^n_{T,d}=0.040, f^n_{T,s}=0.037,$ $|n(k)\rangle$ is the neutron state with momentum $k$, and $m_n$ is the neutron mass\,\cite{Lin:2019uvt}. 
The Dirac spinors in Eq.\eqref{eq:amp} are approximated in the non-relativistic limit as, $u(k)\approx \sqrt{2m_n} ~(\xi_s ~\xi_s)^T$, where $\sum_s \xi_s \xi^\dagger_s =I$. Moreover, $t-m^2_\phi \rightarrow -m^2_\phi$, since $t \approx \mathcal{O}(v^2)$ is velocity-suppressed for non-relativistic DM particles. The spin-averaged matrix amplitude of the interaction is
\begin{align}
    \overline{|\mathcal{M}|^2} =900 \left(\frac{\mu \lambda_f}{\rm{GeV}}\right)^2 \left(\frac{m_n}{m_\phi}\right)^4, 
\end{align}
which goes into the DM-neutron elastic scattering cross-section, that determines the DM capture rate inside the NS,
\begin{align}
    \sigma_{\chi n} = 7.1\times 10^{-45} ~{\rm cm^2} \left(\frac{m_n}{m_\phi}\right)^4 \left(\frac{\mu}{m_\chi+m_n}\right)^2 \left(\frac{\lambda_f}{10^{-9}}\right)^2.
\end{align}
It is apparent that one can achieve geometric limit of DM-neutron scattering for a range of DM masses with suitable choices of $\mu$ and $\lambda_f$, apparent from Fig.\,\ref{fig:model}. 

\subsection{Annihilation cross-section of dark matter \& freeze-in production}
In this section, we calculate the thermal average of the captured DM annihilation cross-section for the process, $\chi +\chi \rightarrow \bar{f}+f$ inside the NS,  as follows:
\begin{align}
    \sigma & = \frac{1}{8\pi}\frac{\mu^2 \lambda_f^2}{(s-m_\phi^2)^2+\Gamma^2_\phi m^2_\phi} \frac{1}{\sqrt{(1-4m^2_\chi/s)}}(1-4m^2_f/s)^{3/2},
    \label{eq:cross}
\end{align}
where
\begin{align}
    \Gamma_\phi &= \frac{\lambda^2_f m_\phi}{8\pi}\sqrt{1-\frac{4 m^2_f}{m^2_\phi}}+ \frac{\mu^2}{32\pi m_\phi}\sqrt{1-\frac{4m^2_\chi}{m^2_\phi}}\,.
\end{align}
DM annihilation to light quarks is a s-channel process, in which resonance on-shell production of $\phi$ is possible, if the center of mass energy is $s\simeq m^2_\phi$ and $m_\phi > 2m_\chi$.
For $\Gamma_\phi \ll m_\phi$, the cross-section can be written in the following form, exploiting the narrow-width approximation,
\begin{align}
    \sigma = \frac{\mu^2\lambda_f^2}{8\pi} \frac{(1-4m^2_f/s)^{3/2}}{\sqrt{1-4m^2_\chi/s}}\left[\frac{\pi}{\Gamma_\phi m_\phi} \delta(s-m^2_\phi)+ {\rm P} \left(\frac{1}{(s-m^2_\phi)^2}\right)\right],
\end{align}
where P stands for the principal value. The second term in the above expression is the off-shell contribution of the DM annihilation. 
The thermal average of the resonant part of the cross-section can be found analytically as \cite{Gondolo:1990dk},
\begin{align}
    \expval{\sigma v}_{\rm res} = \frac{\lambda^2_f}{64\pi m^2_\chi}\frac{\mu^2 m^2_\phi}{\Gamma_\phi m^3_\chi }\left(1-\frac{4m^2_f}{m^2_\phi}\right)^{3/2}\left(1-\frac{4m^2_\chi}{m^2_\phi}\right)^{1/2} \frac{\pi x K_1(xm_\phi/m_\chi)}{K^2_2(x)} ,
    \label{eq:res}
\end{align}
where $x=m_\chi/T$ with $T$ being the temperature of the thermal bath in the early Universe. The thermal average of the off-shell part cannot be calculated analytically in a closed form, i.e.,
\begin{align}
  \expval{\sigma v}_{\rm{off}}= \frac{\lambda^2_f}{64\pi m^2_\chi} \frac{x~\mu^2 }{ m^3_\chi K^2_2(x)}\, {\rm P}\left(\int \frac{(1-4m^2_f/s)^{3/2}}{(s-m^2_\phi)^2\sqrt{1-4m^2_\chi/s}} ~ s (s-4m^2_\chi)^{1/2} K_1(x \sqrt{s}/m_\chi) ds \right)\,.
\end{align}
However, for non-relativistic DM with $m_\chi/T\gtrsim 3$,  we can approximate both of the above expressions to write total thermal average $\expval{\sigma v}_{\rm th}=\expval{\sigma v}_{\rm res}+\expval{\sigma v}_{\rm{off}}$ as 
\begin{align}\label{eq:xsection}
    \expval{\sigma v}_{\rm th}=\frac{\lambda^2_f}{64\pi m^2_\chi}\left[\frac{\mu^2 m^2_\phi}{\Gamma_\phi m^3_\chi}\left(1-\frac{4m^2_\chi}{m_\phi}\right)^{1/2}\sqrt{\frac{ m_\chi}{2m_\phi}}~(\pi x)^{3/2}~ e^{-x\left(\frac{m_\phi-2m_\chi}{m_\chi}\right)}+\frac{\mu^2}{m^2_\chi(1-m^2_\phi/4m^2_\chi)^2}\right],
    \end{align}
where we have ignored light quark masses. Note that in  Eq.\eqref{eq:xsection}, only the second term survives in the low temperature limit, i.e. $x\gg 1$, giving rise to the s-wave nature of the annihilation cross-section.
%Note that the second term in the above expression, stemming from the off-shell contribution of the cross-section, survives at low temperatures, i.e. $x\gg 1$, indicating s-wave behavior of the annihilation cross-section.
The s-wave cross-section is crucial for the enhanced annihilation to occur from the BEC as discussed in the previous section. Because the typical velocity of the DM particles in a BEC is even smaller than that in a thermal core, p-wave annihilation cross-section would be highly suppressed and will not lead to the enhanced annihilation scenario considered in this work. Assuming $m_\phi = 2m_\chi (1+\delta)$, where $\delta<1$, the s-wave contribution of DM annihilation cross-section becomes
\begin{align}
    \expval{\sigma v}_{\rm th} \simeq \frac{1.5\times 10^{-38}{~\rm cm^3 ~s^{-1}}}{\delta^2}  \left(\frac{\lambda_f}{10^{-9}}\right)^2 \left(\frac{{\rm GeV}}{m_\chi}\right)^4 \left(\frac{\mu}{{\rm GeV}}\right)^2.
    \label{eq:sigma1}
\end{align}

Since $\mu$ is a dimensionful coupling, we can also have $\Gamma_\phi \gg m_\phi$ for $\mu \gg m_\phi$, for which the narrow-width approximation fails.  In this case, with similar hierarchy of DM and mediator masses as before, the cross-section takes the form
\begin{align}
\sigma &\approx \frac{\mu^2 \lambda^2_f}{8\pi \Gamma^2_\phi m^2_\phi}(1-4m^2_\chi/s)^{-1/2} = \frac{64 \pi \lambda^2_f}{\mu^2}\frac{1+\delta}{\sqrt{2\delta}}(1-4m^2_\chi/s)^{-1/2},
\label{eq:sigma2}
\end{align}
and its thermal average reads
\begin{align}
    \expval{\sigma v}_{\text{th}}&\approx  \frac{64\pi \lambda_f^2}{\mu^2} \frac{1+\delta}{\sqrt{2\delta}} = 2.4\times 10^{-39}  {\rm cm^3 ~s^{-1}}~\frac{1+\delta}{\sqrt{2\delta}}\left(\frac{10^3 ~{\rm GeV}}{\mu}\right)^2 \left(\frac{\lambda_f}{10^{-9}}\right)^2.
\end{align}
As opposed to the former case, the thermal average of DM annihilation is independent of both DM and the mediator masses and completely controlled by $\mu$ and $\lambda_f$. Therefore, DM annihilation cross section is velocity-independent in the nonrelativistic limit, as apparent from Eq.(\ref{eq:sigma1}) and Eq.(\ref{eq:sigma2}). This is important for the viability of the phenomenological results discussed in the previous section.

It is worth noting that NS heating due to DM annihilation is sensitive to the late-time (low temperature) value of the thermally averaged DM annihilation cross-section, which can also be responsible for DM production in the early universe. With such a small annihilation rate, the production of DM particles in the early universe can happen via \textit{the freeze-in} mechanism \cite{Hall:2009bx,Yaguna:2011qn,Elahi:2014fsa,Bernal:2017kxu}. In the present scenario, DM particles is produced from the SM thermal bath via same annihilation process, but in the reverse direction, i.e. $\bar{f}  +f\rightarrow \chi +\chi$. The co-moving number density of DM is given by \cite{Hall:2009bx,Elahi:2014fsa},
\begin{align}
    Y_\chi \approx 1.32 ~g^{1/2}_* m_\chi M_P \int^\infty_{x_R} \frac{\expval{\sigma v}_{\rm th}}{x^2} Y^{eq^2}_{\chi} (x) dx,
    \label{eq:yield}
\end{align}
where $x_R$ corresponds to the reheating temperature, i.e., $x_R=m_\chi/T_R$, $g_*$ is the effective degrees of freedom of the thermal bath, $Y^{eq}_\chi (x)= 0.11 g^{-1}_* x^2 K_2(x)$ is the co-moving equilibrium number density of DM with $K_n$ being the modified Bessel function of second kind, and $\expval{\sigma v}_{\rm th}$ is the thermal average of DM annihilation cross-section defined in Eq.(\ref{eq:cross}), which we compute following Ref.\,\cite{Gondolo:1990dk}. Here although the process is actually the reverse of DM annihilation, the DM yield $Y_\chi$ depends on the DM annihilation cross-section, defined in Eq.(\ref{eq:cross}). This is due to the energy conservation and the CP- symmetry ($|\mathcal{M}|^2_{\chi\chi\rightarrow \bar{f}f}=|\mathcal{M}|^2_{\bar{f}f\rightarrow \chi \chi}$) of the annihilation process, which is exploited in the Boltzmann equation for the DM yield.    
\begin{figure}[t]
    \centering
    \includegraphics[scale=1.0]{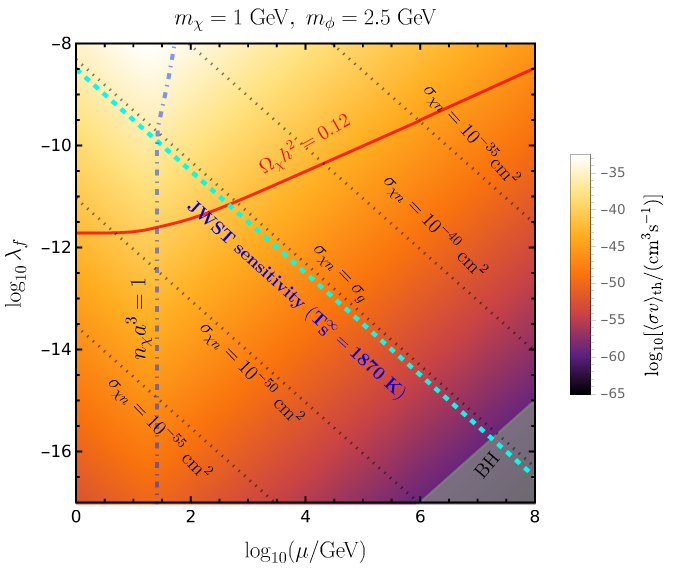}
\caption{\textit{A freeze-in dark matter example with BEC formation inside NS} : Colormap of the thermally averaged DM annihilation cross-section, $\langle\sigma v\rangle$ in the parameter plane of $\lambda_f$ and $\mu$. The DM-neutron cross-section, $\sigma_{\chi n}$ is also shown by the black dotted lines. Here the DM and mediator masses are fixed at $m_\chi=1\,{\rm GeV}$ and $m_\phi=2.5\,$GeV. The red line shows the observed DM relic density. The region above the cyan dashed line is observable by JWST due to enhanced DM annihilation. The gray-shaded region  is excluded by the observation of $10^{10}$ year old NS, which is not transmuted to a BH due to excessive DM capture. In the region to the left of the blue dot-dashed line, the ideal gas approximation holds for DM in the condensate.}
\label{fig:model}
\end{figure}

To understand the relationship between DM relic abundance and various parameters of the model, we present a semi-analytic estimate of the relic density below.
With $\Gamma_\phi \ll m_\phi$, at high temperatures, $\expval{\sigma v}_{\rm th}$ is well approximated by $\expval{\sigma v}_{\rm res}$ as defined in Eq.(\ref{eq:res}). DM freeze-in production starts from $T=T_R$ to $T=m_\phi/3$, beyond which $\expval{\sigma v}_{\rm res}$ drops exponentially, rendering DM production inefficient. Hence, most of the freeze-in production takes place in the relativistic regime $m_\chi/T \ll 1$. Hence, we can perform the integration in Eq.(\ref{eq:yield}) analytically, using an approximate form of the Bessel function\footnote{$K_n (x) \sim  \frac{1}{2} \Gamma(n) (\frac{x}{2})^{-n}$ for $x\rightarrow 0$ and $n>0$ \cite{stegun}.} appearing in thermally averaged cross-section and the equilibrium number density.  The final co-moving DM density is can be written as 
\begin{align}
    Y_\chi(x=3m_\chi/m_\phi) = 4.3\times 10^{-10} \left(\frac{100}{g_*}\right)^{3/2}\left(\frac{\lambda_f}{1.6\times10^{-12}}\right)^2 \left(\frac{{\rm GeV}}{m_\phi}\right).
    \label{eq:DMreso}
\end{align}
Notably, the DM abundance is insensitive to the trilinear coupling $\mu$ and the reheating temperature $T_R$. This is a consequence of the narrow-width resonance, where $\Gamma_\phi$ is dominated by the channel, $\phi \rightarrow \chi \chi$. This is because $\lambda_f$ is taken to be sufficiently small, i.e. $\lambda_f \ll \mu/ m_\phi$ in our analysis. 
 
On the other hand, when $\Gamma_\phi \gg m_\phi$ ( for $\mu \gg m_\phi$), the cross-section is approximated in the high temperature limit as
\begin{align}
    \sigma \approx \frac{\mu^2 \lambda^2}{8\pi} \frac{1}{16T^4+\dfrac{\mu^4}{1024 \pi^2}},
\end{align}
where we have put $s=4T^2$ since $T\gg m_\phi, m_\chi$. Clearly, for $T> \mu/8\sqrt{2} $, the annihilation cross-section is suppressed as the temperature increases, resulting in reduced DM production. Most of DM production takes place when $m_\phi\ll T \lesssim \mu/8\sqrt{2}$, for which the thermal average becomes independent of temperature, $\expval{\sigma v} \simeq 128 \pi \lambda^2/\mu^2$. As a result, the final DM abundance in this case can be obtained as
\begin{align}
    Y_\chi \approx 4.3 \times 10^{-10} \left(\frac{100}{g_*}\right)^{3/2}\left(\frac{\lambda_f}{1.2\times 10^{-10}}\right)^2 \left(\frac{{10^5~\rm GeV}}{\mu}\right). 
    \label{eq:DMnon}
\end{align}
Unlike in the previous case, the DM abundance depends on the trilinear coupling $\mu$, via the lower limit, i.e. $x=8\sqrt{2}m_\chi/\mu$ to $x\approx m_\phi$ of integration in Eq.(\ref{eq:yield}). In this case also, the DM abundance is independent of the reheating temperature, as long as $T_R > \mu$. In the present scenario, we always assume $T_R$ to be higher than all other energy scales.
%, i.e. $T_R > \mu_{\text{max}}$ for our exact numerical results shown in Fig.\ref{fig:model}.

Now, we can write the combined result for DM density $\Omega_\chi h^2$ using Eq.(\ref{eq:DMreso}) and Eq.(\ref{eq:DMnon}) in the two cases as follows,
\begin{align}
\frac{\Omega_\chi h^2}{0.12} =
\begin{cases}
     \left(\dfrac{m_\chi}{{\rm GeV}}\right) \left(\dfrac{100}{g_*}\right)^{3/2}\left(\dfrac{\lambda_f}{1.6\times10^{-12}}\right)^2 \left(\dfrac{{\rm GeV}}{m_\phi}\right)~~ \text{for}~ \Gamma_\phi\ll m_\phi,\\
     \left(\dfrac{m_\chi}{{\rm GeV}}\right)\left(\dfrac{100}{g_*}\right)^{3/2}\left(\dfrac{\lambda_f}{1.2\times 10^{-10}}\right)^2 \left(\dfrac{{10^5~\rm GeV}}{\mu}\right)~~\text{for}~ \Gamma_\phi\gg m_\phi.
\end{cases}
\label{eq:relic}
\end{align}
We have normalized the density with the value $\Omega_\chi h^2=0.12$ measured by the Planck 2018 experiment\,\cite{Planck:2018vyg}. 
In Fig.\,\ref{fig:model}, we show a heatmap of the thermally averaged DM annihilation cross section in the parameter plane of $\mu$-$\lambda_f$. 
The solid red line shows the line for $\Omega_\chi h^2 =0.12$.
The resonant process dominates the annihilation for $\mu \lesssim 10$ GeV which explains the lighter shaded region in the upper-left part of Fig.\,\ref{fig:model}. The cross section decreases for $\mu\gtrsim 10$ GeV toward the right of the plot. This is also reflected in the shape of the $\Omega_\chi h^2 =0.12$ line.

We also show the lines of constant $\sigma_{\chi n}$ in Fig.\,\ref{fig:model} in relation to the NS capture phenomenology discussed in the previous section. In the entire region, $\expval{\sigma v} \lesssim 10^{-40}~{\rm cm^3 ~ s^{-1}}$, for which $1$ GeV mass DM forms the BEC, as already shown in the previous section. The shaded region below the light gray line is excluded from the existence of the $10^{10}$ year old NS in the DM environment with $\rho_\chi=0.4\,{\rm GeV ~cm^{-3}}$. In this region with  $\expval{\sigma v} \lesssim \mathcal{O}(10^{-60}){\,\rm cm^3 ~s^{-1}}$, a NS can transmute into a BH due to excessive capture of DM particles. The parameter region above the cyan dashed line can be probed using the JWST observation of `heated' NS due to enhanced DM annihilation. As the observed DM relic abundance line also falls in this region, JWST possess a unique capability to test this model in the near future.

In Fig.\,8, we show the result with ideal BEC approximation, i.e., without any self-interaction. In presence of DM self-interaction, the parameter space related to the NS observables may change, as already discussed before. In the next section, we examine the validity of this assumption by estimating DM self-scattering within the present DM model.
%within this current framework and check the most conservative parameter space where our results remain reliable.}

\subsection{Self-scattering cross-section of dark matter}
DM particles can undergo self-scattering via the trilinear interaction with the scalar mediator. The self-scattering amplitude of DM receives contributions from $s,t,u$ channel diagrams, as follows :
\begin{align}
    \mathcal{M}_{\chi\chi\rightarrow \chi\chi} = \mu^2 \left[\frac{1}{4m^2_\chi-m^2_\phi+i\Gamma_\phi m_\phi}-\frac{2}{m^2_\phi}\right],
\end{align}
where we have put $s=4m^2_\chi$, $t=u=0$ for non-relativistic DM scattering. For $\mu_\phi >m_\phi$, the DM self-scattering goes into the non-perturbative regime, signifying a strong self-interaction. As we already discussed in Sec.\ref{sec:self}, an accurate treatment of self-interaction in the BEC state warrants a detailed study using GP equations, thereby here we only here we stipulate our discussion for the perturbative region. Therefore, the DM self-scattering cross-section is given by
\begin{align}
    \sigma_{\chi\chi} = \frac{(\mu/m_\phi)^4}{128\pi m^2_\chi}\left[\frac{m^4_\phi}{(4m^2_\chi-m^2_\phi)^2+\Gamma^2_\phi m^2_\phi}+4-\frac{4m^2_\phi(4m^2_\chi-m^2_\phi)}{(4m^2_\chi-m^2_\phi)^2+\Gamma^2_\phi m^2_\phi}\right],
\end{align}
for $\mu/m_\phi \leq 4\pi$. Now, similar to the previous section, we calculate the self-scattering length and compare it with the typical inter-particle separation of DM condensate. In Fig.\,\ref{fig:model}, we show the contour of $n_\chi a^3 =1$ by blue dot-dashed line, to the right of which the system enters the strong self-interaction regime, where an involved analysis is required including effects of DM capture and annihilation. We calculate the DM number density assuming the minimum radius of the dark core in the condensate state. Therefore, the $n_\chi a^3 = 1$ line  represents a very conservative limit on the validity of our results.

\section{Conclusion \& Future Outlook}
\label{sec:discussion}
In this work, we have demonstrated the implications of Bose-Einstein condensate formation by bosonic DM particles captured inside a neutron star. We have shown that DM capture inside a NS can constrain \emph{freeze-in} models of Bosonic DM with small scattering cross sections that are beyond the reach of even future-generation direct detection experiments. 

We first study the evolution of the NS temperature in the presence of DM annihilation to examine the onset of the BEC transition for a given set of dark sector parameters. Subsequently, we have shown that the BEC formation without any self-interaction in the DM core will drastically enhance their annihilation rate due to the sudden contraction of the dark core inside the NS increasing the number density. In particular, the capture-annihilation equilibrium can now be achieved with extremely small DM annihilation cross-sections, leading to an elevated surface temperature of old NS within the projected sensitivity of JWST. The effect of BEC formation was not taken into account in previous studies\,\cite{Garani:2020wge,Saha:2025fgu}. Therefore, our work shows that a much larger DM parameter space can now be probed by ongoing and future observations of JWST. More importantly, the enhanced annihilation scenario will generically probe DM-nucleon cross sections that are orders-of-magnitude smaller than what even the next-generation direct detection experiments can reach\,\cite{XLZD:2024nsu,SuperCDMS:2022kse,Angloher:2025fzw}.

First, we have shown the impact of the enhanced DM annihilation for the s-wave cross section to the total DM annihilation cross-section and the DM-neutron elastic cross-section within a model-independent framework, without considering DM formation in the early Universe. Afterward, we realize this scenario in a particle physics model comprising a scalar DM candidate and a comparable mass scalar mediator, where the DM-SM couplings are determined by interactions of the mediator with DM through a trilinear coupling and with SM quarks via a Yukawa coupling. With suitable choice of parameters, we have obtained the DM-neutron elastic cross-section around its geometric limit, while the annihilation cross-section is several orders of magnitude smaller than the canonical WIMP annihilation cross section. In this framework, the DM relic density is set by the freeze-in mechanism in the early universe. Hence, the present scenario is consistent with early universe observation, and can be probed by near-future NS observation. To the best of our knowledge, \emph{this is the first study of a freeze-in model in the context of DM capture inside the NS}.  

We have given a projected sensitivity of the JWST to probe this model using its infra-red frequency band observation of old NS. However, looking for old NSs in a wide region of the sky remains challenging, as the JWST is optimized for deep observations rather than wide sky coverage\,\cite{Gardner:2006ky,Bramante:2021dyx,Raj:2024kjq}. To this end, ground-based Rubin/LSST  observations can effectively complement JWST, due to its large sky coverage and its ability to identify population of old NSs\,\cite{LSST:2008ijt,Bramante:2021dyx}. In addition to JWST, upcoming facilities such as the Extremely Large Telescope\,\cite{ELT} and the Thirty Meter Telescope\,\cite{TMT} will be capable of measuring NS surface temperatures down to $\sim 10^3$\,K using advanced imaging instruments\,\cite{Baryakhtar:2017dbj,Raj:2024kjq}. Consequently, the combination of deep-observation telescopes with wide sky-survey facilities will be able to observe more old NSs and enable us to study their thermal evolution.

We have also briefly discussed how our scenario is modified in presence of self-interaction. However, a quantitative study of effects of DM self-interaction is rather involved and kept for future study. This stems from the fact the DM population is time-dependent inside a BEC due to capture and annihilation of DM particles. In particular, DM annihilation can play a dominant part in determining the spatial extent and overall stability of the condensate due to high particle density compared to the non-BEC case. We plan to explore this phenomenology in a future work.

\section*{Acknowledgment}
We thank  Sohini Pal for designing the schematic diagram presented in Fig.\ref{fig:illus}. We gratefully acknowledge useful discussions with Sulagna Bhattacharya, Ranjan Laha, Ranjini Mondol, Nirmal Raj, Tarak Nath Maity, Anupam Ray, and Akash Kumar Saha. AD and DG acknowledge financial support from the Department of Atomic Energy through the projects RSI 4001 and RSI 4009. AD acknowledges ANRF for the financial support through PMECRG (Grant no.~ANRF/ECRG/2025/001012/PMS).

\bibliography{refs}
\bibliographystyle{jhep}
%%%%%%%%
\end{document}